\documentclass[reprint,aps,pra,superscriptaddress,nofootinbib]{revtex4-2}
\usepackage[utf8]{inputenc} 
\usepackage[T1]{fontenc}    
\usepackage{hyperref}       
\usepackage{url}            
\usepackage{booktabs}       
\usepackage{amsfonts}       
\usepackage{nicefrac}       
\usepackage{microtype}      
\usepackage{xcolor}         
\usepackage{graphicx}
\usepackage{longtable} 
\usepackage{array}     
\usepackage{amsmath}
\usepackage{amssymb}        
\usepackage{textcomp}
\usepackage[final]{fixme}
\fxsetup{layout=inline, theme=color}

\usepackage{siunitx} 
\usepackage{subcaption} 
\usepackage{soul} 
\usepackage[nameinlink,noabbrev]{cleveref}

\makeatletter
\newsavebox\pandoc@box
\newcommand*\pandocbounded[1]{
  \sbox\pandoc@box{#1}%
  \Gscale@div\@tempa{\textheight}{\dimexpr\ht\pandoc@box+\dp\pandoc@box\relax}%
  \Gscale@div\@tempb{\linewidth}{\wd\pandoc@box}%
  \ifdim\@tempb\p@<\@tempa\p@\let\@tempa\@tempb\fi
  \ifdim\@tempa\p@<\p@\scalebox{\@tempa}{\usebox\pandoc@box}%
  \else\usebox{\pandoc@box}%
  \fi%
}
\makeatother
\makeatletter
\def\maxwidth{\ifdim\Gin@nat@width>\linewidth\linewidth\else\Gin@nat@width\fi}
\def\maxheight{\ifdim\Gin@nat@height>\textheight\textheight\else\Gin@nat@height\fi}
\makeatother
\setkeys{Gin}{width=\maxwidth,height=\maxheight,keepaspectratio}
\makeatletter
\def\fps@figure{htbp}
\makeatother

\NewDocumentCommand\citeproctext{}{}

\makeatletter
\let\@cite@ofmt\@firstofone
\def\@biblabel#1{}
\def\@cite#1#2{{#1\if@tempswa , #2\fi}}
\makeatother
\newlength{\cslhangindent}
\newlength{\csllabelwidth}
\newenvironment{CSLReferences}[2] 
{\begin{list}{}{%
	\setlength{\itemindent}{0pt}
	\setlength{\leftmargin}{0pt}
	\setlength{\parsep}{0pt}
	\ifodd #1
	\setlength{\leftmargin}{\cslhangindent}
	\setlength{\itemindent}{-1\cslhangindent}
	\fi
	\setlength{\itemsep}{#2\baselineskip}}}
{\end{list}}
\usepackage{calc}

\newcommand{\CSLLeftMargin}[1]{\parbox[t]{\csllabelwidth}{\strut#1\strut}}
\newcommand{\CSLRightInline}[1]{\parbox[t]{\linewidth - \csllabelwidth}{\strut#1\strut}}

\newenvironment{widemath}{}{}

\expandafter\let\csname longtable*\endcsname\undefined
\expandafter\let\csname endlongtable*\endcsname\undefined
\makeatletter
\def\ptc@save#1{\expandafter\let\csname ptc@\string#1\endcsname#1}
\def\ptc@restore#1{\expandafter\let\expandafter#1\csname ptc@\string#1\endcsname}
\def\ptc@arraycmds#1{#1\@array#1\@@array#1\@tabular#1\@tabarray#1\array
  #1\endarray#1\endtabular#1\@mkpream#1\@classx#1\@arrayacol#1\@tabacol
  #1\insert@column#1\@arraycr#1\@xarraycr#1\@xargarraycr#1\@yargarraycr}
\ptc@arraycmds\ptc@save
\AddToHook{begindocument/end}{\ptc@arraycmds\ptc@restore
  \expandafter\let\csname endtabular*\endcsname\endtabular}
\makeatother
\AtBeginDocument{\renewcommand\bibnumfmt[1]{}}
\hypersetup{pdftitle={On the scaling limits of hypermultiplexed photonic tensor cores},
  pdfauthor={Hugh Morison, Nayem Al-Kayed, A. Aadhi, Bhavin Shastri}}
\makeatletter
\let\ptc@addcontentsline\addcontentsline
\def\ptc@title{title}
\def\addcontentsline#1#2#3{\def\ptc@tmp{#2}%
  \ifx\ptc@tmp\ptc@title\else\ptc@addcontentsline{#1}{#2}{#3}\fi}
\makeatother
\renewenvironment{widemath}{\begin{widetext}}{\end{widetext}}

\begin{document}
\title{On the scaling limits of hypermultiplexed photonic tensor cores}
\author{Hugh Morison}
\email{hugh.morison@queensu.ca}
\affiliation{Department of Physics, Engineering Physics, and
Astronomy, Queen's University}
\author{Nayem Al-Kayed}
\email{nayemal.kayed@queensu.ca}
\affiliation{Department of Physics, Engineering Physics, and
Astronomy, Queen's University}
\author{A. Aadhi}
\email{aadhi@iitd.ac.in}
\affiliation{Optics and Photonics Centre, Indian Institute of Technology
Delhi}
\author{Bhavin Shastri}
\email{shastri@ieee.org}
\affiliation{Department of Physics, Engineering Physics, and
Astronomy, Queen's University}
\begin{abstract}
Hypermultiplexed photonic tensor cores compute matrix--tensor products
making use of spatial, temporal, and spectral degrees of freedom to
increase throughput. Published projections reach hundreds of TOPS by
increasing the number of wavelengths \(N\), the spatial fan-out \(S\),
and the symbol rate \(f_B\). We model this architecture class, including
many sources of receiver noise, spectral crosstalk between channels, an
amplified link budget with saturation limits, and the power consumption
of every subcomponent of the analog core. All operating points in the
design space have a power-independent resolution ceiling related to
their maximum signal-to-noise ratio. Spectral crosstalk sets this
ceiling with a bound \(N f_B \le 1.1\) THz for 4 bits and
\(N f_B \le 0.79\) THz for 8 bits in the C-band. With current
components, however, the amplified link is the binding constraint;
amplifier saturation limits the power that reaches the detectors, so
receiver noise sets the resolution up to 4 bits, and amplifier noise
becomes dominant at higher resolutions. For a thin-film lithium niobate
architecture with quantum-dot amplifiers, the most efficient designs
give 3.9, 2.5, and 1.4 TOPS/W at 2, 3, and 4 bits, with fan-out factors
of 2 to 8. We find optima in the spatial fan-out and the symbol rate,
but the design space is remarkably flat over channel counts from 10 to
64. Coupling loss and the saturation power per Watt for the amplifiers,
rather than more wavelengths or faster symbol rate, are the highest
leverage changes for efficiency.
\end{abstract}
\maketitle

\section{Introduction}\label{sec:introduction}

Integrated analog photonic computing architectures have emerged as
promising candidates for high throughput matrix and tensor
multiplication operations, with the potential to reduce system power
consumption and end-to-end latency compared to their digital
counterparts, primarily single-instruction multiple-data (SIMD)
processors (e.g.~graphics processing units (GPUs)). Many photonic matrix
multiplication architectures have been proposed, with some of the first
prominent integrated architectures based on Mach-Zehnder interferometer
(MZI) meshes {[}1{]} and microring resonator (MRR) weight banks {[}2{]}.
Other notable architectures have been achieved using photonic crossbar
arrays {[}3, 4{]}, or comb sources in conjunction with dispersive delay
lines to achieve convolutions {[}5{]}. Dong et al.~added radio-frequency
subcarriers as a third multiplexing dimension, next to wavelength and
space, in a crossbar with non-volatile phase-change weights {[}6{]}.
Recent commercial investments have yielded large scale photonic matrix
multiplication accelerators which demonstrate photonic compute
co-packaged with CMOS control and data converters, and they have
reported whole system power on real workloads {[}7, 8{]}. These systems
have high precision, leverage spatial parallelism, and operate at symbol
rates of a few GHz, but they use neither wavelength parallelism nor
optical amplification in service of the computation.

Many proposals and demonstrations in the photonic computing literature
identify and leverage one or several favourable properties of computing
with optics---high bandwidth optical components, easy spatial or
spectral parallelism, or passive routing and low energy switching---and
then report a metric such as throughput in operations per second (OPS),
or compute efficiency in joules per operation or equivalently operations
per second per Watt (OPS/W), but a common criticism is that there isn't
consensus on the boundary on which sources of power consumption are
counted {[}9, 10{]}. As noted in reviews on this subject, an optical
processor will most likely need to combine several of these favourable
properties in order to outperform electronic processors {[}11, 12{]}.
Many of the favourable properties share the same physical origin, the
large bandwidth of the optical spectrum, but the best manner of
leveraging all of these benefits simultaneously is still an open problem
{[}11{]}. Reviews also note that optical loss, amplification, and the
power consumption of data converters at the boundary of the analog data
path all limit the precision and the efficiency of large systems {[}10,
13, 14{]}. Less attention has been given to the interdependence of these
properties when they draw on the same finite resources (the finite
optical bandwidth, limited source power, or the gain of shared optical
amplifiers), and efforts to exploit several simultaneously run into more
constraints than unbounded scaling arguments might suggest.

The tension in these scaling arguments is especially clear in
wavelength-division-multiplexed (WDM) architectures that try to exploit
both high spectral channel count, and a high per-channel modulation
bandwidth. Scaling both of these variables is advantageous, but with a
finite spectrum these two benefits are in direct conflict with one
another. As more optical channels are added, the bandwidth that can be
allocated to each channel before they begin to overlap must shrink, or
else the resulting crosstalk imposes a cap on resolution of the
computation. To many practitioners in the field, this tradeoff may be
obvious enough to not even bear mentioning, since there are standard
mitigation strategies such as guard bands and pulse shaping to
efficiently use the spectrum. Previous analyses have mitigated this
conflict with a hard rule, either a channel count limit set by the
usable spectrum divided by the per-channel rate {[}15{]}, or a resonance
width tied to the sampling rate {[}16{]}. After accounting for these
mitigations, many published scaling arguments for the throughput \(T\)
treat the number of spectral channels \(N\) and the modulation symbol
rate \(f_B\) as if they were freely combinable {[}3, 17, 18{]}, but a
full analysis incorporating all scaling parameters into measures of the
computing performance may have implications for the design of
subcomponents in optical computing systems (WDM
multiplexers/demultiplexers/routers, and optical amplifiers), since
optical computing systems have different design criteria and tradeoffs
from WDM optical communications. To our knowledge, no analysis carries
the result of spectral crosstalk into the signal-to-noise ratio, where
it sets a limit on \(N f_B\) that depends on the target resolution.
Recent WDM hypermultiplexed demonstrations that target high channel
counts train their optical layers at 3-bit precision with
quantization-aware training, against a measured noise-equivalent
resolution of 3.5 to 4.5 bits that falls with the symbol rate {[}19,
20{]}. Nonlinearity, measurement bandwidth, noise, and operating
conditions can all set this precision, which motivates a quantitative
account of the error sources at scale. Scaling in wavelength and in
space also splits the optical power among more channels and paths, and
the optical gain that restores this power adds amplified spontaneous
emission, which sets a second limit on precision {[}4, 14{]}.

This previous demonstration by Pappas et al.~is an example of a
\emph{hypermultiplexed} photonic tensor core (PTC) architecture, which
leverages cascaded broadband MZI modulators (MZMs) and an arrayed
waveguide grating router (AWGR) to simultaneously exploit the spatial
and spectral degrees of freedom {[}19, 20{]}. The literature also calls
this class of architecture multidimensional {[}21{]} or hyperdimensional
{[}20{]}, but we prefer the term hypermultiplexed, after Ou et al., who
also multiplex in space, time and wavelength {[}18{]}, to avoid
confusion with the hyperdimensional computing paradigm. The initial
proposal demonstrated 20 GBaud operation with a 16 \(\times\) 16 AWGR
for a projected throughput of 163.8 TOPS {[}19, 21{]}, subsequently
extended with a microcomb source and a 32-port AWGR, operated as an
effective 16 \(\times\) 16 system with 16 wavelength lines at 32 GBaud,
for a projected parallel throughput of \(2 \cdot 16^3 \cdot 32\) GBaud
\(\approx 262\) TOPS {[}20{]}. Both throughput figures assume that all
the paths operate at once, while the experiments drive one weight
modulator and one input modulator, which are switched across the paths.
The closest architectural cousin to the hypermultiplexed architecture,
proposed by Ou et al., encodes the inputs on seven directly modulated
VCSELs on a 100 GHz WDM grid and the weights using seven TFLN
modulators, with a passive power splitter in place of the AWGR {[}18{]}.
It reports 5 to 6 bits of computing accuracy and a projected 0.98 TOPS
at 10 GSa/s, although the experiments drove one channel at 10 GSa/s or
two channels at 100 MS/s at a time. Related work on cascaded MZMs
without spectral parallelism has pushed symbol rates to 60 GBaud using
thin-film lithium niobate (TFLN) {[}22{]}, and beyond 100 GBaud when
embedded as a component in an Ising machine, albeit at reduced bit
precision {[}23{]}. The same product, with one weight row per
wavelength, one broadband modulator for the inputs, a demultiplexer and
time-integrating receivers, was demonstrated by Sludds et al.~as a
protocol that streams weights to edge devices, at kHz rates {[}24{]}.
Given these different demonstrations all push on different axes of the
design space, either by pushing the data rate as fast as possible with
one channel or by increasing the number of wavelengths and modulators
while maintaining a moderate data rate, the hypermultiplexed PTC is one
of the best case studies we can use to examine the tradeoffs explicitly.
The follow-ups to the original proposal estimated energy efficiency from
component power draws in the literature and an insertion-loss budget
that sets the number and gain of the optical amplifiers {[}19, 20{]},
and the 262 TOPS work bounds the crosstalk of its router from the
measured precision {[}20{]}. They did not, however, model the noise of
the amplified link, or optimize jointly over channel count, symbol rate,
amplification and resulting resolution, which requires the explicit
accounting of relationships between power consumption, signal-to-noise
ratio, and effective bit resolution.

Nahmias et al.~derived the energy per multiply--accumulate (MAC)
operation of photonic processors from shot noise, converter energy and
amplifier power, and bounded the aggregate rate of a WDM processor by
the usable optical bandwidth {[}15{]}. An analysis put forward by Tait
derived analytic scaling laws for the power consumption of MRR
weight-bank and MZI-mesh networks {[}25{]}, where each power
contribution was defined as a power law in the channel count \(N\) and
the operation frequency \(f\), and the noise-limited laser power was
found to grow exponentially with the resolution \(B\), defined from the
spurious-free dynamic range; this analysis included an RIN limit that
constrains bandwidth at a given resolution, but it excluded optical
amplification and interchannel crosstalk. Al-Qadasi et al.~combined an
optical link budget, a receiver noise model, and the power of lasers,
drivers, thermal tuning, ADCs, memory interfaces, and SOAs for silicon
MZI-mesh and MRR accelerators at 1 to 10 GSa/s {[}26{]}, and found the
largest matrix that the available laser power supports at each
resolution, with crosstalk as a fixed penalty and SOAs as fixed-gain
stages without saturation. Brückerhoff-Plückelmann et al.~estimated the
efficiency and area density of MRR, crossbar and incoherent MZI-mesh
processors with full converter interfaces, at one accuracy target
equivalent to 8-bit inputs and outputs with about 4-bit weights
{[}16{]}; in that study, the only spectral constraint applies to the MRR
banks, where each resonance must be about ten times wider than the
sampling rate, so the free spectral range imposes a cap on the product
\(N f_{Sa}\), and inter-channel crosstalk does not enter the noise
budget. None of these analyses models the interchannel crosstalk as a
noise term that grows with \(N f_B\), the saturation of shared optical
amplifiers, or a joint optimization of the scale parameters at a fixed
resolution.

In this work we close this gap for hypermultiplexed PTCs by developing a
unified framework that incorporates computational throughput, the power
of the modelled electronic and photonic components
(Sec.~\ref{sec:power_model}), a link budget with amplifier placement,
and the noise of the optical link and receiver, including the
spectral-crosstalk term that couples \(N\) and \(f_B\). For each set
precision level, we optimize the spectral channel count \(N\), the
spatial parallelism \(S\), the operational symbol rate \(f_B\), the
optical source power, and the optical amplifier placement strategy
jointly for minimum power consumption. The framework gives three
power-independent limits on resolution and shows where each one is the
limiting constraint on the architecture. Spectral crosstalk bounds
\(N f_B\) at about 1.1 THz for 4 bits in the C-band, the laser source
RIN bounds \(f_B\), and the in-band crosstalk of the router, if its
static part is not calibrated out, bounds the resolution of large
designs. With today's components, however, we find that the amplified
link is the limiting factor. Due to the link loss when scaling such
architectures, the saturation power, power consumption and noise of the
optical amplifiers set both the resolution and the efficiency of the
best designs. The compute efficiency (TOPS/W) does not grow
monotonically with \(N\), \(S\) or \(f_B\). For a heterogeneous
thin-film lithium niobate platform with quantum-dot amplifiers at about
1.1 W per stage, the most efficient designs give 3.9, 2.5, and 1.4
TOPS/W at 2, 3, and 4 bits. These designs use 30 to 35 wavelengths, a
spatial fan-out that falls from 8 at 2 bits to 2 at 4 bits, and symbol
rates that fall from 34 GBaud at 2 bits to 25 GBaud at 4 bits. Each of
these designs uses a single amplifier stage, and the receiver noise sets
their resolution. From 5 bits onward, a second amplification stage is
necessary, and signal--spontaneous noise sets the resolution. We also
find that the compute efficiency is almost flat over most of the design
space. At 2 to 4 bits, every channel count from about 9 to 64 and all
symbol rates spanning around an order of magnitude come within 20\% of
the best design. The channel count can therefore be chosen for
throughput at little cost in compute efficiency. Crosstalk is at most
13\% of the noise of the most efficient designs, and its share is
largest at low precisions around 2 and 3 bits. The goal of a system
architect should therefore not be ever more wavelength channels or
modulation speed, but lower coupling losses and better amplifiers
(efficiency, noise figure and saturation power), which would allow
additional channels to be beneficial.

\begin{figure*}
\centering
\pandocbounded{\includegraphics[keepaspectratio,alt={Hypermultiplexed Tensor Core Schematic. Schematic of the photonic tensor core (PTC) architecture proposed by Pappas et al.~The architecture uses broadband Mach-Zehnder Modulators (MZMs) and an Arrayed Waveguide Grating Router (AWGR) to achieve highly parallel matrix-tensor multiplication, leveraging the spatial, spectral, and temporal degrees of freedom. A weight matrix is encoded with the first set of MZMs, while a tensor of input vectors is encoded onto the second set of MZMs, performing an analog optical multiplication via their electro-optic transmission. This schematic shows all peripheral components that contribute to the power consumption and noise of the compute core. Colours in this diagram correspond to optical wavelengths, with colour gradients indicating wavelength-division multiplexed components.}]{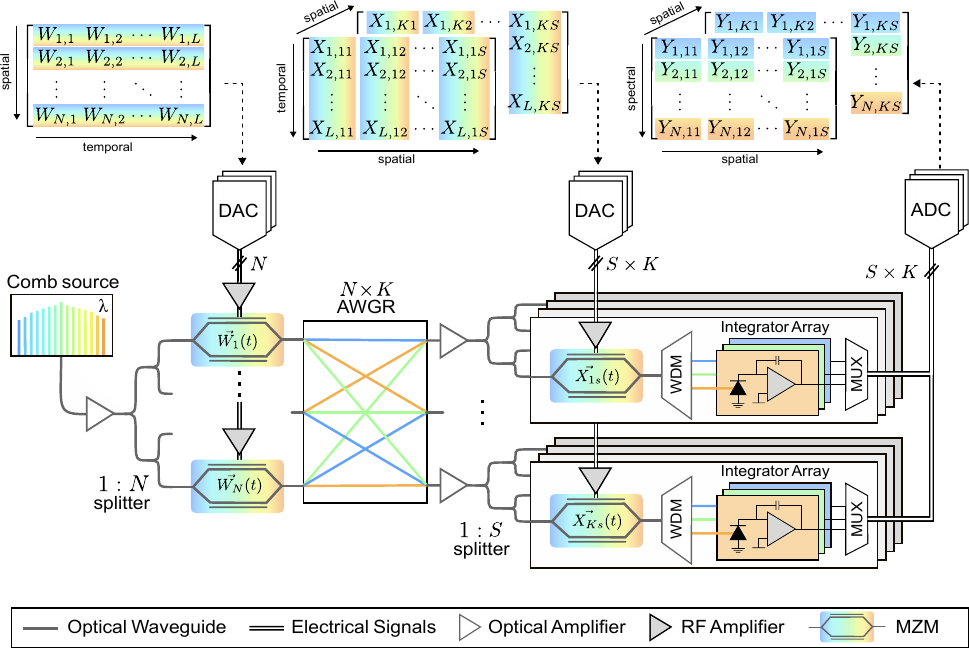}}
\caption{Hypermultiplexed Tensor Core Schematic. Schematic of the
photonic tensor core (PTC) architecture proposed by Pappas et al.~The
architecture uses broadband Mach-Zehnder Modulators (MZMs) and an
Arrayed Waveguide Grating Router (AWGR) to achieve highly parallel
matrix-tensor multiplication, leveraging the spatial, spectral, and
temporal degrees of freedom. A weight matrix is encoded with the first
set of MZMs, while a tensor of input vectors is encoded onto the second
set of MZMs, performing an analog optical multiplication via their
electro-optic transmission. This schematic shows all peripheral
components that contribute to the power consumption and noise of the
compute core. Colours in this diagram correspond to optical wavelengths,
with colour gradients indicating wavelength-division multiplexed
components.}\label{fig:tensor_core_schematic}
\end{figure*}

\subsection{The Hypermultiplexed Architecture}\label{sec:architecture}

The hypermultiplexed photonic tensor core {[}21{]} depicted in
Fig.~\ref{fig:tensor_core_schematic} uses cascaded intensity modulators
and an AWGR to encode an analog matrix-tensor multiplication across
three physical degrees of freedom---wavelength, space, and time. This
section describes the architecture put forward in their work, identifies
the scaling parameters that the rest of the paper analyzes, and
introduces the throughput scaling law that the remaining analysis will
constrain.

The architecture is described by four scaling parameters, each mapping
onto a distinct physical dimension. The number of wavelength channels
\(N\) is also the number of modulators at the first stage and the number
of AWGR input ports, and sets the scale in the spectral degree of
freedom. The number of AWGR output ports is \(K\), where \(K \leq N\).
Each output port receives all \(N\) weights per row (one per wavelength)
so \(K\) sets how many spatial copies of the weight matrix are fed to
the subsequent stage. We take \(K=N\) throughout, primarily for
simplicity, but also because using less spatial fan-out is strictly
inefficient from the computational perspective. After the AWGR, there is
a further spatial fan-out by a factor \(S\), which allows the system to
multiply the number of independent input vectors processed in parallel.
The symbol rate \(f_B\) of the modulators sets the use of the temporal
degree of freedom by defining the number of computations processed by
the PTC in a given amount of time. With \(K=N\), these degrees of
freedom result in three free parameters describing the hardware scaling,
\(N\), \(S\), and \(f_B\).

The accumulation depth \(L\) is the number of time-domain symbols
accumulated at the receiver before readout, and differs per workload
processed on the PTC. At the readout side, the PTC computes a
matrix-tensor product \(Y = W \cdot X\) between an \(N \times L\) matrix
\(W\) and an \(L \times S \times K\) tensor \(X\) (\(SK\) input vectors
of length \(L\)), producing an \(N \times S \times K\) tensor \(Y\).
\(L\) is therefore the inner dimension of the product, and can be chosen
at runtime. The numerical results in this work use square weight
matrices, \(L=N\). \(L\) affects neither the throughput nor the optical
link budget; it enters the analysis only for the power consumption of
the data converters (Sec.~\ref{sec:power_model}). Using a larger inner
dimension and accumulating more symbols before readout reduces the ADC
power as \(\sqrt{N/L}\).

In this architecture, a multi-wavelength source (an \(N\)-line frequency
comb, or \(N\) wavelength-multiplexed lasers) feeds an \(N\)-way fan-out
into the first bank of MZMs. Each of these MZMs in the first bank
encodes one row of \(W\) at \(f_B\) on all \(N\) wavelengths. The \(N\)
waveguides then enter the AWGR, which passively routes each wavelength
of each input port into a different output port. Each AWGR output is
further split \(S\) ways, producing \(SK\) physical paths in total,
which are input to the second bank of MZMs. Each of the MZMs in the
second bank then encodes one input vector onto all \(N\) wavelengths,
which performs the analog product between the elements of the input
vector and the elements of the weight matrix carried on the signals
input to the MZM bank. A demultiplexer after each MZM then separates the
wavelengths onto \(N\) receivers (photodetectors paired with
transimpedance integrators) which accumulate the product over the \(L\)
symbols, where the accumulated products are then digitized.

The distinguishing component of this architecture which enables
hypermultiplexed computing is the AWGR. Its cyclic routing property
distributes wavelength \(\lambda_i\) arriving at input port \(u\) to
output port \(v = 1 + ((i + u - 2) \bmod N)\). Concretely, for
\(N = 4\), input port 1 sends \(\lambda_1\) to output 1, \(\lambda_2\)
to output 2, and so on, and input port 2 sends \(\lambda_1\) to output
2, \(\lambda_2\) to output 3, etc. Output port 1 then receives
\(\lambda_1\) from input 1, \(\lambda_4\) from input 2, \(\lambda_3\)
from input 3, and \(\lambda_2\) from input 4. Each output port therefore
collects one wavelength from every input port, each of which carries one
of the \(N\) rows of \(W\). In general, after the second modulator bank,
the photodiode for \(\lambda_i\) on split path \(s\) of output port
\(k\) then integrates the sum \(Y_{u,s,k}=\sum_l W_{u,l} X_{l,s,k}\).
The \(NSK\) photoreceivers output all the elements of \(Y\), each of
which represents one multiplication operation and one addition operation
per symbol, which gives Eq.~\ref{eq:throughput}.

The architecture uses \(N + SK\) modulators to realize the matrix-tensor
product, and, ignoring all physical constraints, has a computational
throughput of
\begin{equation}\protect\phantomsection\label{eq:throughput}{
T = 2 N S K f_B \quad \text{ops/s.}
}\end{equation}

We start our analysis from this scaling relationship. Taken at face
value, Eq.~\ref{eq:throughput} suggests unbounded growth in throughput
by pushing any of \(N\), \(S\), \(K\), or \(f_B\). The remainder of this
work addresses the constraints on this growth, and finds where the most
efficient operating points lie in the design space after taking into
account the system-wide implications of scaling the design parameters.

The first constraint on Eq.~\ref{eq:throughput} is the density of the
spectral packing and crosstalk which couple \(N\) and \(f_B\). Packing
more channels into a finite optical band forces each channel to occupy a
narrower bandwidth, while increasing the symbol rate increases the
bandwidth of each channel. These two goals conflict with one another,
and beyond a certain \(N \cdot f_B\) product, adjacent channels begin to
overlap and the associated crosstalk imposes a certain amount of noise
on every output signal, which imposes a cap on signal-to-noise ratio
(SNR), which is related to the bit resolution of the computation. The
second constraint on Eq.~\ref{eq:throughput} comes from the other noise
sources in the system which impose a limit on useful throughput. The
losses in the system from splitting and device coupling and insertion
create a need for amplification in the link which scales with \(N\) and
\(S\), and the amplifier noise introduced translates into a large
penalty on the achievable bit resolution in the receiver. Increasing the
scale of the system can still increase \(T\), but this is only useful up
to the point where the bit resolution is sufficient for the target
computation. The third constraint on Eq.~\ref{eq:throughput} doesn't
relate to the bit resolution, but the compute efficiency. The power
consumption of all the active devices grows with scale and symbol rate.
Data converters, receivers, driver amplifiers, optical amplifiers, and
thermal stabilization of the optics all consume power that grows with
\(N\), \(S\), \(K\), and \(f_B\), in some cases faster than the photonic
subsystem. Determining which of these elements dominates the power
consumption at the most efficient design points is one of the outcomes
of this analysis (Sec.~\ref{sec:design_space}).

In order to simplify the analysis, we consider two special cases for the
scaling parameters throughout this work. Setting \(S = 1\) removes the
post-AWGR fan-out, reduces the second modulator bank from \(SK\) to
\(K\) modulators, gives a product with \(K\) input vectors,
\(T = 2NKf_B\) and substantially less loss. On the other hand, setting
\(N = S = K\) (symmetric hypermultiplexing) pushes all three physical
degrees of freedom together, which allows a single parameter to describe
the system scale.

For readout, the baseline architecture {[}21{]} uses one ADC per
receiver, \(NSK\) in total, but we introduce one further simplification
here to reduce the ADC count. Because the integrating receiver reads out
once every \(L\) symbols, one ADC can instead be time-multiplexed across
the \(N\) wavelength channels of a spatial output, which reduces the ADC
count to \(SK\) when \(L \ge N\). While this reduces the device count,
it does not affect the aggregate conversion power, because each shared
ADC runs \(N\) times faster (Sec.~\ref{sec:power_model}). The
accumulation depth \(L\) is the variable that amortizes conversion
power.

\section{Results}\label{sec:results}

By incorporating all the sources of noise and power consumption in all
components of the PTC, we develop the model described in
Sec.~\ref{sec:methods} which relates the physical attributes of each
component and the scale parameters. Sec.~\ref{sec:resolution_results}
presents the upper limits to the bit resolution of the system from the
closed-form relationships in Sec.~\ref{sec:receiver_model} and
Sec.~\ref{sec:spectral_model}, without accounting for the link budget of
the system. Sec.~\ref{sec:link_results} adds the link budget, including
consideration of amplifier placement, and gives the maximum resolution
that an amplified photonic link can deliver under typical parameters for
the physical constants of current technology platforms.
Secs.~\ref{sec:design_space}, \ref{sec:frontier} add a model of power
consumption for all components in the system and find the design with
the highest throughput and compute efficiency for each target
resolution. Each candidate design considered in these sections uses
physically plausible parameters and parameter limits, such as integer
numbers of channels, paths, and numbers of amplifiers, as well as
saturation caps on amplifiers, and realistic values for noise figures of
all components (Sec.~\ref{sec:link_budget}). We evaluate two technology
platforms, described in sec.~\ref{sec:instantiation}, a monolithic
silicon platform with heterogeneous III--V amplifiers, and a thin-film
lithium niobate platform with silicon nitride (SiN) passives with
attached quantum-dot amplifiers. The gain of each stage can be tuned
anywhere below its maximum, with a proportional saving in power
consumption (Sec.~\ref{sec:power_model}). For each target resolution,
the search finds the design with the lowest total power that reaches the
target.

\subsection{Upper Limits to PTC
Resolution}\label{sec:resolution_results}

\begin{figure*}
\centering
\pandocbounded{\includegraphics[keepaspectratio,alt={Upper limits to PTC resolution. (a) Power-independent resolution ceiling \textbackslash mathrm\{SNR\}\_\textbackslash infty with contributions from crosstalk and RIN, shown in the (f\_B, N) plane under a Gaussian-shaped spectral profile, with iso-ENOB contours at 1--8 bits. The dashed line is the packing limit of an ideally band-limited system (\textbackslash beta\{=\}1). The markers on the plot represent published demonstrations from {[}18--20, 22, 23{]}, placed at their channel count and symbol rate, and labelled with their measured precision converted to the convention in Eq.~. (b) The universal saturation curve of Eq.~ for representative \textbackslash gamma; all operating points in this work lie on this family of curves.}]{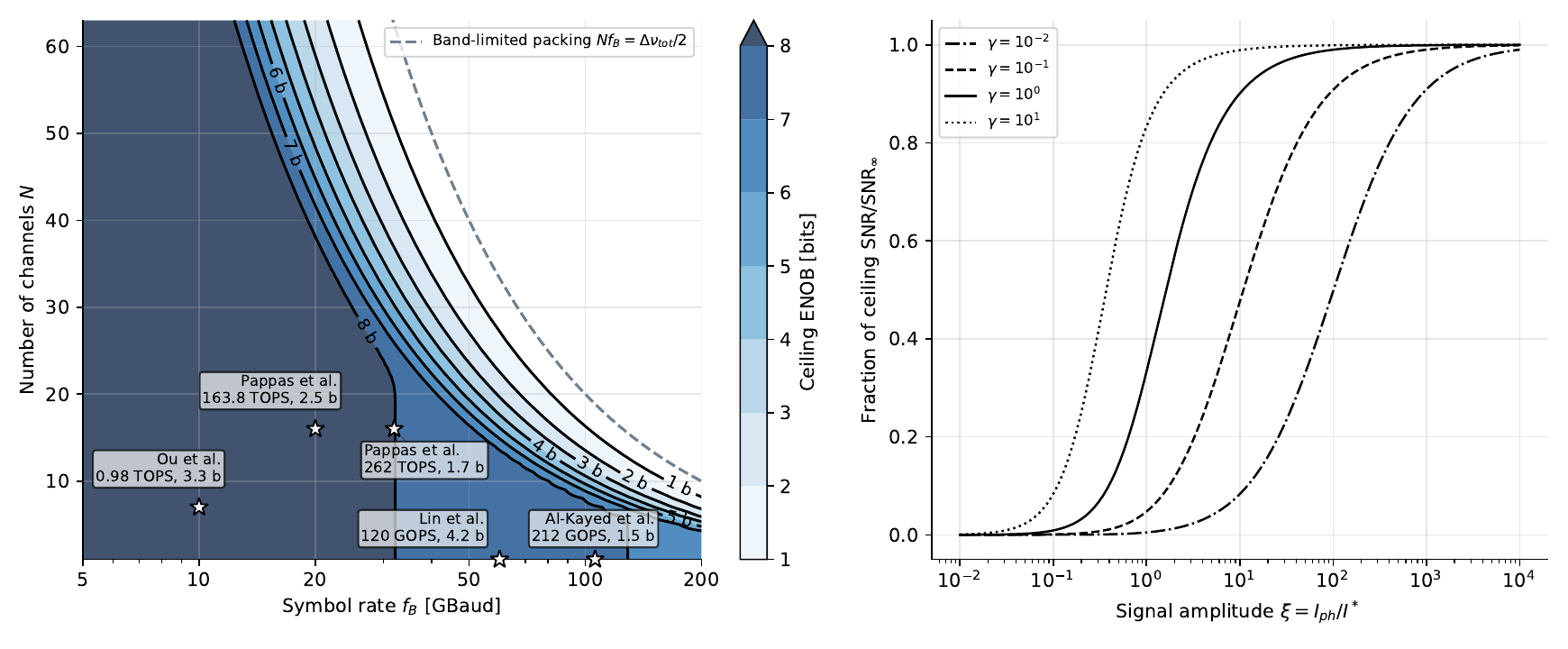}}
\caption{Upper limits to PTC resolution. (a) Power-independent
resolution ceiling \(\mathrm{SNR}_\infty\) with contributions from
crosstalk and RIN, shown in the \((f_B, N)\) plane under a
Gaussian-shaped spectral profile, with iso-ENOB contours at 1--8 bits.
The dashed line is the packing limit of an ideally band-limited system
(\(\beta{=}1\)). The markers on the plot represent published
demonstrations from {[}18--20, 22, 23{]}, placed at their channel count
and symbol rate, and labelled with their measured precision converted to
the convention in Eq.~\ref{eq:enob}. (b) The universal saturation curve
of Eq.~\ref{eq:snr_ceiling} for representative \(\gamma\); all operating
points in this work lie on this family of
curves.}\label{fig:wall_master}
\end{figure*}

Regardless of the details of the optical link, every MAC operation
performed in this architecture results in a signal at the receiver
represented by the photocurrent \(I_{ph}\), which must compete with the
total noise current. The photocurrent carries the product of two
signals, which are modulated onto optical carriers with
intensity-modulation index \(m\) (Eq.~\ref{eq:cascaded_mzm}). For
operands drawn independently with per-operand variance \(\sigma_{op}^2\)
(and a variance of their product \(\sigma_d^2\)), the signal-to-noise
ratio of a single MAC takes the form,
\begin{equation}\protect\phantomsection\label{eq:master_snr}{
\mathrm{SNR} = \frac{\sigma_d^2\, m^4\, I_{ph}^2}{A_0 + A_1 I_{ph} + A_2 I_{ph}^2},
}\end{equation} wherein every independent source of noise in the system
has been grouped into the \(A_i\) depending on how they scale with
\(I_{ph}\).

The terms which are independent of the signal (the receiver's thermal
noise and the spontaneous--spontaneous beat noise), are combined into
\(A_0\) and represent a floor on the total noise. The noise terms in
\(A_1\) (shot noise and signal--spontaneous beat noise from optical
amplification) are linear in the signal current and therefore cause the
SNR to scale with the square root of the signal power. The terms
captured in \(A_2\) (laser relative intensity noise, RIN, and
interchannel crosstalk) scale with the signal power itself, and
represent a ceiling on the achievable SNR. The exact expressions for the
\(A_i\) are detailed in Sec.~\ref{sec:methods}. This type of
classification is standard in analog microwave photonics {[}27, 28{]};
the aspect that is specific to this class of optical computing
architectures is the data statistics \(\sigma_d^2 m^4\) in the
numerator, and a crosstalk contribution to \(A_2\) set by the modulation
statistics of the neighbouring channels. Normalizing by the current
\(I^\ast = A_0/A_1\), where the linear noise group equals the constant
group, collapses Eq.~\ref{eq:master_snr} to a universal two-parameter
form,
\begin{equation}\protect\phantomsection\label{eq:snr_ceiling}{\mathrm{SNR} = \mathrm{SNR}_\infty\,\frac{\gamma \xi^2}{1 + \xi + \gamma \xi^2},
}\end{equation} where \(\xi = I_{ph}/I^\ast\) is the normalized signal
strength, and \(\gamma = A_2 A_0/A_1^2\) is a measure of how quickly
increasing signal strength approaches the maximum signal-to-noise ratio
\(\mathrm{SNR}_\infty = \sigma_d^2 m^4/A_2\). Eq.~\ref{eq:snr_ceiling}
is plotted in Fig.~\ref{fig:wall_master}(b). Every point in the design
space of this class of architecture, indeed any WDM intensity-modulated,
direct-detection (IM/DD) computing link on any platform, lives on this
curve family, with \(\gamma\) set by the specifics of the hardware. The
first notable property of this relationship is that
\(\mathrm{SNR}_\infty\) is a power-independent ceiling on any platform's
achievable SNR under a specified set of operating conditions. At this
point, because the noise power scales proportionally with the signal
power, no amount of laser power or amplifier gain will yield more
resolution at this maximum. Second, the single shape parameter
\(\gamma\), which is also set by specific platform parameters, measures
how much optical power must be expended to reach the maximum SNR. This
relationship only reaches the ceiling \(\mathrm{SNR}_\infty\)
asymptotically, so there are diminishing returns to the system
resolution past a certain level of optical power. We assume throughout
this work that we remain under the threshold for nonlinear optical
distortion, but including these effects would contribute a fourth term
to the denominator of Eq.~\ref{eq:master_snr} and cause the SNR to start
decreasing with further power.

In the computational context, we care how the SNR of the system
translates to an effective number of bits (ENOB). We quantify ENOB using
the definition of IEEE Std 1241 {[}29{]}, referenced to the full range
of the product signal \(I_\text{FS} = 2 m^2 I_{ph}\),
\begin{equation}\protect\phantomsection\label{eq:enob}{
\begin{aligned}
B &= \log_2\!\left(\frac{I_\text{FS}}{\sqrt{12}\,\sigma_n}\right)
= \log_2\!\left(\frac{m^2 I_{ph}}{\sqrt{3}\,\sigma_n}\right) \\
&= \frac{\mathrm{SNR}_{dB} - 10\log_{10}(3\sigma_d^2)}{6.02},
\end{aligned}
}\end{equation} where \(\sigma_n\) is the square root of the noise
variance accounting for all sources of noise. An ideal quantizer with
step \(\Delta\) has an RMS error of \(\Delta/\sqrt{12}\), so the ENOB is
the bit width of an ideal quantizer, spanning the product range, whose
RMS error equals the noise of the analog computation. Published
demonstrations of analog computing usually report a noise-equivalent bit
resolution, \(NEB = \log_2(I_\text{FS}/\sigma_n)\) {[}15, 30{]}. For the
same RMS error, the NEB is \(\log_2\sqrt{12} = 1.79\) bits above the
ENOB, and where we compare with published systems we give both. We use
the ENOB as an approximate point of comparison with \(b\)-bit digital
arithmetic (see the caveats in Sec.~\ref{sec:frontier}).

Fig.~\ref{fig:wall_master}(a) maps the equivalent ENOB of the ceiling
\(\mathrm{SNR}_\infty\) over the \((f_B, N)\) plane for a
\(\Delta\nu_{tot} = 4~\text{THz}\) comb corresponding to the optical
C-band, using the definitions of the noise parameters \(A_i\) in
Sec.~\ref{sec:methods}. The two noise components of \(A_2\),
interchannel crosstalk and RIN noise, contribute to the ceiling on the
bit resolution. Interchannel crosstalk, which comes from the beat
between each channel's carrier and the modulation sidebands of its
neighbours, and from the detected leaked intensity of the neighbours
(see Sec.~\ref{sec:methods}), produces the near-hyperbolic contours
where \(N f_B\) is approximately constant. Requiring the crosstalk
ceiling to reach a target resolution \(B^*\), with
\(\mathrm{SNR}^* = 3\sigma_d^2\, 4^{B^*}\) from Eq.~\ref{eq:enob},
yields, \begin{equation}\protect\phantomsection\label{eq:wall}{
N f_B \;\le\; \frac{\Delta\nu_{tot}}{c_\text{eff}\,
\sqrt{2 \ln (2) B^*  + \ln\!\big(6 k_\text{eff}/m^4\big)}},
}\end{equation} where \(k_\text{eff}\) is an effective coefficient which
describes the strength of all the neighbours' contributions to the
spectral crosstalk, and \(c_\text{eff}\) is the effective bandwidth of
the signal, the demultiplexer, and the receiver in units of \(f_B\),
both of which are fitted to the full crosstalk model (see
Sec.~\ref{sec:methods}). Using parameters achievable on technology today
(Tbl.~\ref{tbl:hardware_params} and Tbl.~\ref{tbl:shared_params}), this
bound sits at \((N f_B)_{max} = 1.35/1.06/0.90/0.79~\text{THz}\) for
2/4/6/8 bits, respectively. Because the target enters only through
\(\sqrt{B^* + \text{const}}\), the bound is remarkably tolerant to
changes in precision, so that upgrading from 4 to 8 bits (16× the number
of noise-resolvable levels) costs only about 25\% of the \(N f_B\)
product. In addition to the cross-talk limitation, the relative
intensity noise (RIN) of the laser source imposes a second limit to the
SNR for high symbol rates, independently of \(N\). From
Eq.~\ref{eq:enob} and the noise bandwidth \(\Delta f = f_B / 2\), the
RIN term alone allows a resolution \(B\) up to the symbol rate
\(f_B \le 2m^4/(3 r\, 4^{B})\), with \(r = 10^{RIN/10}\). A source with
a typical \(RIN = -155\) dB/Hz caps the resolution at 8 bits at about 32
GBaud and 7 bits at about 130 GBaud, visible in
Fig.~\ref{fig:wall_master}(a) where the contours deviate from their
hyperbolae into vertical segments.

Eq.~\ref{eq:wall} bounds the spectral occupancy \(N f_B\) at a given
resolution, but different points along the hyperbola can have different
throughput and efficiency metrics. With \(K = N\),
Eq.~\ref{eq:throughput} gives \(T = 2NS (N f_B)\), so gains to
throughput are readily made by trading \(N\) for \(f_B\) (at a fixed
\(S\), doubling \(N\) while halving \(f_B\) results in a doubling of
\(T\)). If we denote the right-hand side of Eq.~\ref{eq:wall} as
\(C(B)\), the spectrally limited throughput is \(T \le 2NS\,C(B)\). The
spectral limit therefore leaves the split between \(N\) and \(f_B\)
open, and we must incorporate the link budget and the power accounting
to decide.

For algebraic simplicity, the contours of Fig.~\ref{fig:wall_master}(a),
which show the maximum achievable bit resolution, are computed using
signal pulse shapes with a Gaussian power-spectral-distribution (PSD),
approximating weakly shaped signalling. Unshaped NRZ pulses have more
slowly decaying spectral tails (Sec.~\ref{sec:spectral_model}), so for
NRZ these contours are optimistic. One natural objection to this part of
the analysis is that spectral pulse shaping should remove the effect of
cross-talk on the bit precision, since band-limited pulses pack to
\(\Delta\nu_{ch} = (1+\beta)f_B\) with vanishing linear crosstalk. This
scenario is included in the plot as the dashed line, drawn at
\(\beta{=}1\), but we must note that shaping is not free here, and the
particular costs of pulse shaping are specific to analog computing. The
first point is that band-limited waveforms overshoot their symbol
levels, typically by a few dB, and confining the modulator drive signal
to the linear regime costs modulation depth \(m\). In the cascaded
modulator PTC scheme, the product result pays for this penalty as
\(m^4\) against the RIN, receiver noise, and amplifier noise, whereas a
single-modulator link would only incur a penalty of \(m^2\).
Additionally, for shaped pulses, the integrating receiver recovers the
exact sum of products only if the pulse shape is orthogonal to its own
shifted copy, which is not true in general, and the deviation would
present another source of undesired noise. Another important cost to
consider is that pulse shaping adds per-lane digital signal processing
(DSP) on \((N + SN)\) transmitters. Each modulator's waveform feeds on
the order of \(N\) MACs per symbol, but a pulse shaping filter with
\(N_{tap}\) taps executes about \(N_{tap}\) digital MAC operations per
symbol. Filtering therefore adds about \(N_{tap}/N\) digital operations
for every analog one, and the DACs must also operate at two or more
samples per symbol, which can increase the power consumption. In a
communications link, the filter output is transported data, so this
power consumption cost is just a part of the transciever, but in the
analog computing context, it competes directly with the analog
operations the shaping is in support of, and is only small when
\(N \gg N_{tap}\) or when the energy of a filtering MAC is negligible
compared to the energy of a general MAC on a competing SIMD processor.
The final consideration against pulse shaping is that any realistic
pulse shaping still leaves a finite residual floor in the stopband,
which will anyway impose a crosstalk ceiling of the same form as
Eq.~\ref{eq:wall}, just with steeper decay in the tails. We defer
computing the exact shape of this ceiling for a realistically
pulse-shaped family in favour of developing an algebraic relationship
that can be readily applied to achieve cross-talk estimates which are
conservative relative to shaped cross-talk (optimistic relative to NRZ),
although numerical solutions with realistic pulse-shaping can easily be
incorporated into our framework.

Fig.~\ref{fig:wall_master}(a) also places published demonstrations of
cascaded-modulator computing on the heatmap {[}18--20, 22, 23{]}. All
five marker labels report the equivalent ENOB for each work, which is
1.79 bits below their reported NEB. The markers are only illustrative
since each published work have small implementation differences that
make an exact direct comparison on a single plot impossible. For
instance, the map assumes \(\Delta\nu_{ch} = \Delta\nu_{tot}/N\), not
the real spacing of each system. The two single-wavelength systems have
no spectral packing, so for them the map shows only how much room the
ceiling leaves for more wavelengths at their symbol rates. The system of
Ou et al.~uses seven lines on a 100 GHz grid at 10 GSa/s, where the
crosstalk ceiling is far above its reported precision. For the Pappas et
al.~system, at the reported 100 GHz line spacing, the crosstalk ceiling
is 2.4 bits at 32 GBaud and 7.5 bits at 20 GBaud. The reported precision
corresponds to an ENOB of 1.7 and 2.5 bits (3.5 and 4.3 bit NEB) at
these symbol rates {[}20{]}, so both lie below the ceiling.

\subsection{Link-Limited Bit Resolution}\label{sec:link_results}

\begin{figure*}
\centering
\pandocbounded{\includegraphics[keepaspectratio,alt={Link-limited resolution at f\_B = 32 GBaud. (a) Maximum achievable ENOB against the number of channels N for the silicon (SiPh) and TFLN scenarios of Tbl.~, with S = 1 (solid) and S = N (dashed). Results come from a search over designs using discrete amplifier stages with variable gain and power saturation caps. The grey line is the power-independent ceiling of Fig.~. (b--e) Noise composition of the design with the highest ENOB for each platform, with S = 1 (b, c) and S = N (d, e).}]{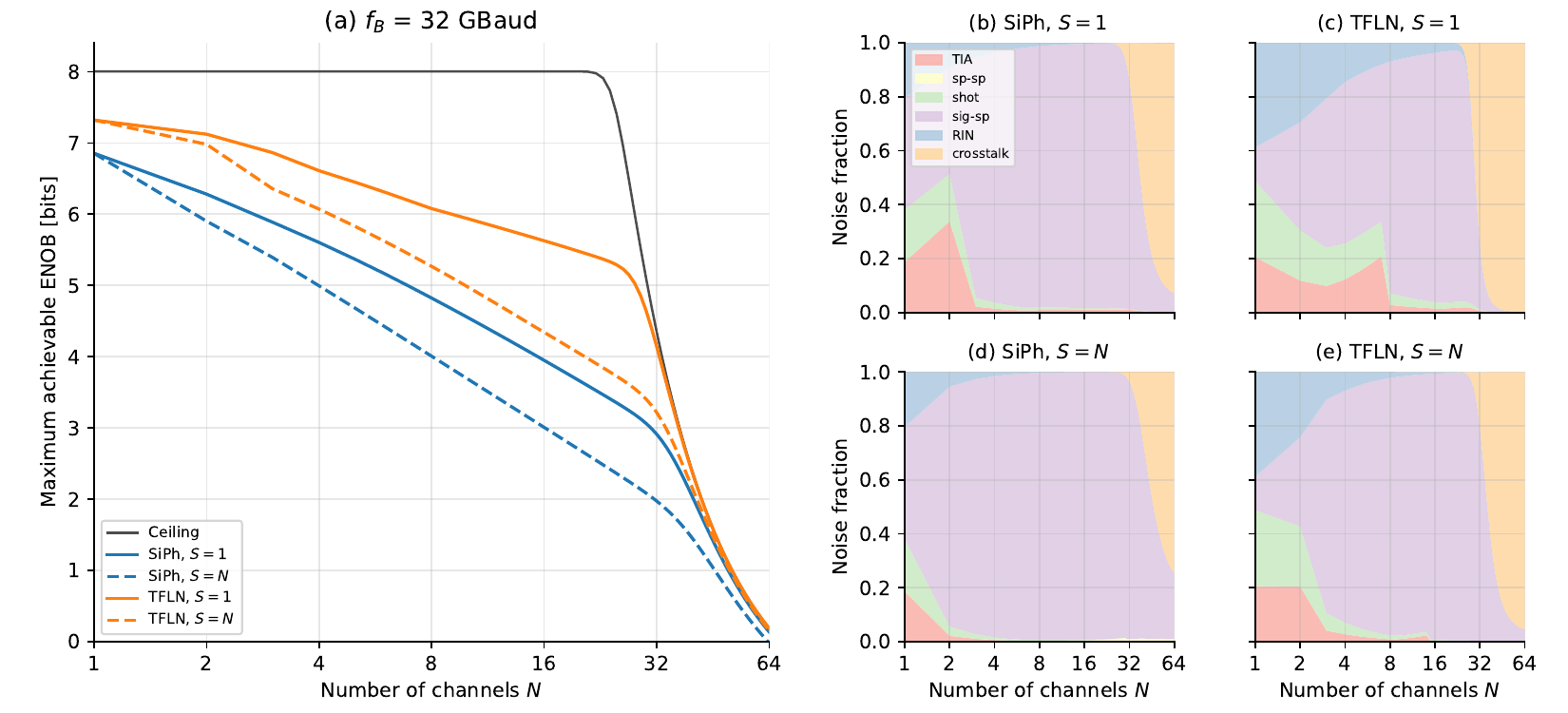}}
\caption{Link-limited resolution at \(f_B = 32\) GBaud. (a) Maximum
achievable ENOB against the number of channels \(N\) for the silicon
(SiPh) and TFLN scenarios of Tbl.~\ref{tbl:hardware_params}, with
\(S = 1\) (solid) and \(S = N\) (dashed). Results come from a search
over designs using discrete amplifier stages with variable gain and
power saturation caps. The grey line is the power-independent ceiling of
Fig.~\ref{fig:wall_master}. (b--e) Noise composition of the design with
the highest ENOB for each platform, with \(S = 1\) (b, c) and \(S = N\)
(d, e).}\label{fig:link_limited}
\end{figure*}

The ceiling represented by Fig.~\ref{fig:wall_master} is an upper limit,
but says nothing about what is achievable for specific choices of our
scale parameters. Reaching the ceiling requires delivering enough
optical power to the detectors, and this architecture must first pay a
broadcast loss that grows as \(10\log_{10}(NS)\) due to the
architecture's inherent fan-out factor, plus 17--20 dB of fixed
insertion and coupling losses due to the modulators and AWGR (23 dB for
the multi-chip TFLN scenario, see the parameters in
Tbl.~\ref{tbl:hardware_params}). At scale, we must therefore introduce
optical amplification to compensate for the substantial loss, and each
stage of required amplification contributes its own noise term through
amplified spontaneous emission of spectral density \(S_{ASE}\). In this
regime, with a power \(P_{det}\) reaching the receiver,
Eq.~\ref{eq:snr_ceiling} reduces to,
\begin{equation}\protect\phantomsection\label{eq:osnr_penalty}{
\mathrm{SNR}_{sig\text{-}sp} = \frac{\sigma_d^2\, m^4}{4}\cdot\frac{P_{det}}{S_{ASE}\, \Delta f},
}\end{equation} which exposes a penalty due to the encoding of the
optical multiplication; the ASE beats against the full optical carrier,
while the useful signal occupies only the fraction \(\sigma_d^2 m^4\) of
the total carrier power (see Sec.~\ref{sec:methods}). Relative to a
single-modulator communications link with the same modulation index and
data statistics, the second modulation incurs a factor
\(m^2 \sigma_{op}^2\) in SNR, which is 4.8 dB for our parameters.
Relative to a full-swing signal, the total penalty is
\(\sigma_d^2 m^4\), or 9.5 dB.

Fig.~\ref{fig:link_limited}(a) shows the maximum achievable bit
resolution as a function of \(N\), with \(S = 1\) and \(S = N\), for
both technology scenarios at 32 GBaud. The results show that realistic
links with realistic amplifiers operate well below the ceiling ENOB. For
\(N \ge 4\) and \(S = 1\), designs using parameters typical of a TFLN
platform sit up to 2.6 bits below it, and with \(S = N\) the gap is up
to 4.1 bits. Below about 30 channels the architecture is therefore
power- and ASE-limited, not ceiling-limited, and the optimal design sits
on the rising slope of the universal curve in
Fig.~\ref{fig:wall_master}(b). Above about 30 channels, the TFLN designs
with \(S = 1\) reach the ceiling, which falls steeply with \(N\) at this
symbol rate. Due to the saturation limit of the amplifiers, the
achievable resolution for all platforms also decreases with increasing
scale. For TFLN with \(S = 1\) it falls from 6.6 bits at \(N = 4\) to
5.6 bits at \(N = 16\), about 0.5 bits per doubling of \(N\), and then
follows the ceiling down to 0.2 bits at \(N = 64\). With \(S = N\) it
falls by about 0.95 bits per doubling from \(N = 4\) to 32. Every shared
amplifier is limited in aggregate output power, so the power available
per channel at any amplification site falls as \(P_{sat}/N\), which, as
the system grows, forces the amplifiers further down the link. Friis'
formula for cascaded noise describes how this limitation degrades the
achievable end-to-end noise performance as \(N\) and \(S\) grow (see
Sec.~\ref{sec:methods}). The SiPh and TFLN cases differ in loss,
amplifier parameters, modulator drive swing and waveguide power cap
together, so the offset between them (up to about 2 bits) cannot be
attributed to any single parameter.

Fig.~\ref{fig:link_limited}(b--e) show the composition of the total
noise of the design with the highest ENOB from different noise sources.
These designs use all the optical power that the limits allow; the
minimum-power designs of Sec.~\ref{sec:design_space} are optimized for
efficiency, so their noise mix moves toward a higher proportion of
receiver noise. Every design requires at least one amplifier stage,
which sits after the first modulator bank for every \(N \ge 2\). For
TFLN with \(S = 1\) and \(N \le 7\), one or two stages before the AWGR
are enough, and the receiver noise, shot noise and RIN remain
significant. From \(N = 8\), a second stage after the second modulator
bank is required, and signal--spontaneous beat noise dominates. Averaged
over \(N = 4\) to 64, the signal--spontaneous noise is 39\% of the noise
for TFLN at \(S = 1\) and 55\% at \(S = N\), and 59--74\% for SiPh.
Crosstalk becomes significant for \(N \gtrsim 30\) at this symbol rate,
where the ceiling itself falls, while RIN dominates only for a single
channel. The prevalence of the signal--spontaneous regime comes from the
saturation limit, which forces the gain downstream, and the
product-encoding penalty of Eq.~\ref{eq:osnr_penalty} raises the
detected power that this regime needs. Below the spectral crosstalk
limit, the biggest lever in improving PTC resolution is therefore by
delivering more optical power and improving the quality of optical
amplifiers (reducing noise figure or increasing the saturation power),
not with more efficient use of the spectrum.

\subsection{Compute Efficiency}\label{sec:design_space}

\begin{figure*}
\centering
\pandocbounded{\includegraphics[keepaspectratio,alt={Design space of the TFLN platform. (a) Heatmap of the compute efficiency \textbackslash eta of the minimum-power designs, in the (f\_B, N) plane, taking the best fan-out S in each cell. Designs under 1 bit are shown in grey. The countour lines show the highest ENOB that each cell can reach at any power value and the blue circles show the most efficient cell for specified target bit resolutions. (b--e) Efficiency \textbackslash eta against spectral occupancy N f\_B at N = 8, 16, 32 and 64, taking the best S for each, for resolution targets of 2 to 5 bits. The dotted vertical lines show the spectral limits of Eq.~. (f) Best efficiency \textbackslash eta\^{}* against N for each target resolution from 2 to 6 bits, taking the best S and f\_B, with circles marking the optimum N\^{}*(B). (g) \textbackslash eta\^{}* against f\_B, taking the best N and S, with circles marking f\_B\^{}*(B). (h) \textbackslash eta\^{}* against S, taking the best N and f\_B, with circles marking S\^{}*(B). All design points in all panels are computed using the fixed resolution target policy of Sec.~.}]{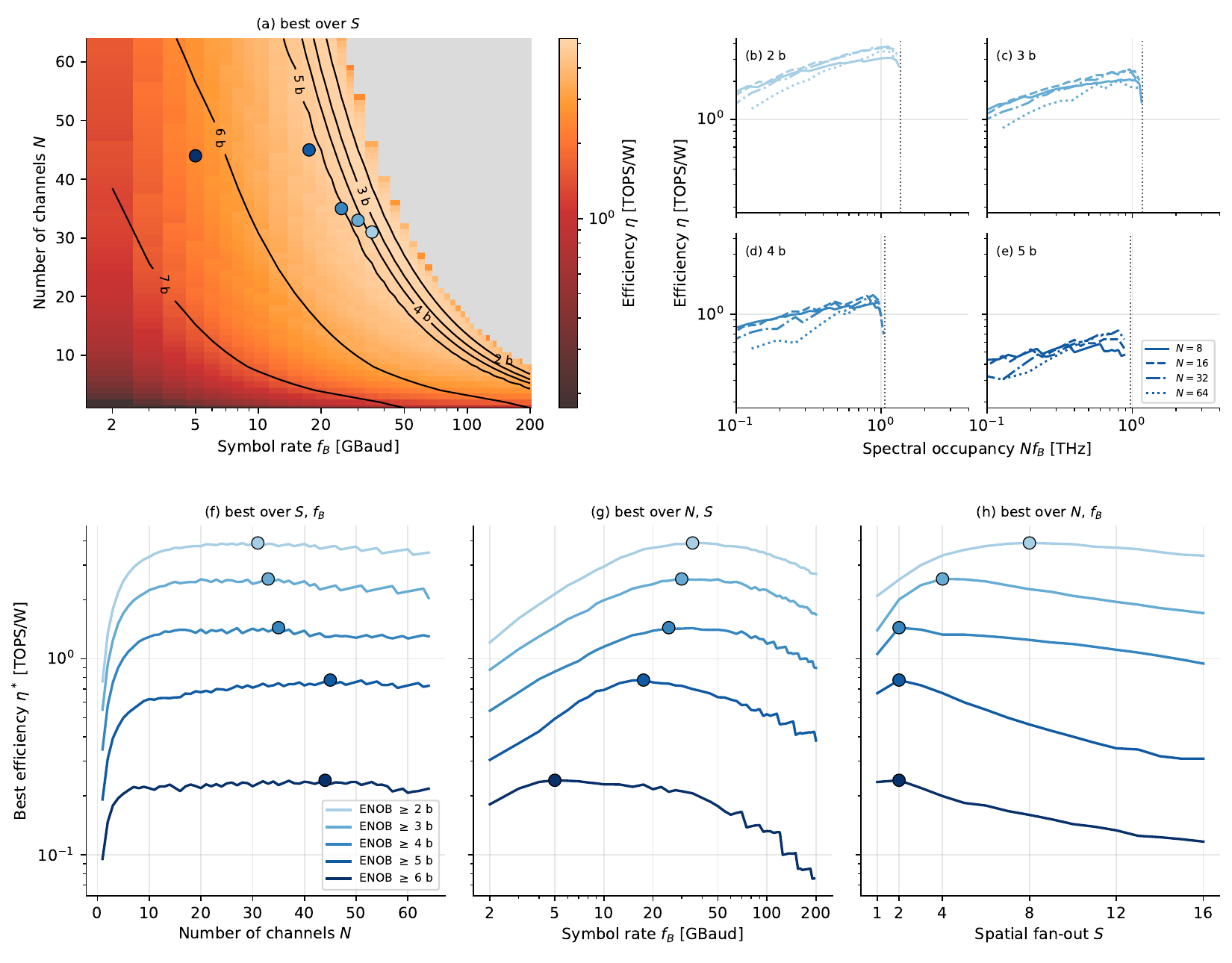}}
\caption{Design space of the TFLN platform. (a) Heatmap of the compute
efficiency \(\eta\) of the minimum-power designs, in the \((f_B, N)\)
plane, taking the best fan-out \(S\) in each cell. Designs under 1 bit
are shown in grey. The countour lines show the highest ENOB that each
cell can reach at any power value and the blue circles show the most
efficient cell for specified target bit resolutions. (b--e) Efficiency
\(\eta\) against spectral occupancy \(N f_B\) at \(N = 8\), 16, 32 and
64, taking the best \(S\) for each, for resolution targets of 2 to 5
bits. The dotted vertical lines show the spectral limits of
Eq.~\ref{eq:wall}. (f) Best efficiency \(\eta^*\) against \(N\) for each
target resolution from 2 to 6 bits, taking the best \(S\) and \(f_B\),
with circles marking the optimum \(N^*(B)\). (g) \(\eta^*\) against
\(f_B\), taking the best \(N\) and \(S\), with circles marking
\(f_B^*(B)\). (h) \(\eta^*\) against \(S\), taking the best \(N\) and
\(f_B\), with circles marking \(S^*(B)\). All design points in all
panels are computed using the fixed resolution target policy of
Sec.~\ref{sec:link_budget}.}\label{fig:design_space}
\end{figure*}

The spectral bound fixes the product \(N f_B\), the spectral occupancy,
available at a given resolution, but it does not choose between many
slow channels and few fast ones, nor the spatial fan-out \(S\). To make
those choices, a specific design objective must be chosen, whether that
is to optimize the number of operations per second or the compute
efficiency, each of which can be evaluated under a fixed resolution. To
evaluate compute efficiency, we account for how the power consumption of
each device scales with \(f_B\), and how their count scales with \(N\)
and \(S\) (see the details in Sec.~\ref{sec:methods}). Some power
consumption terms vary with scale, such as the lasers and first-bank
modulators which scale with \(N\), the input modulators and drivers
which scale with \(SN\), the receivers which scale with \(N^2S\), and
the amplifier count which scales differently depending on their
placement (up to \(N^2 S\)). The power consumption of the modulator
drivers, receiver front ends, and data converters all vary
proportionally to \(f_B\). The data converters scale with both \(N\) and
\(f_B\), and the accumulation depth \(L\) amortizes the power as
\(1/\sqrt{L}\), but the results use square matrices with \(L=N\) for
simplicity (see Sec.~\ref{sec:power_model}).

Fig.~\ref{fig:design_space} maps the efficiency of the minimum-power
designs at each point of the PTC design space (\(N\)--\(S\)--\(f_B\)
space) for the TFLN parameters. Panel (a) shows the efficiency heatmap
over \(N\) and \(f_B\) for all designs with a resolution greater than 1
bit, where each cell is the best design over all values of \(S\),
overlaid with contour lines showing the highest target resolution
achievable at that point. Panels (f--h) show the best efficiency along
each axis, where each panel is optimized over the other two axes, at
specific target resolutions from 2 to 6 bits. The best efficiency
\(\eta^*(B)\) decreases uniformly with the target resolution; for TFLN,
the efficiency decreases from 3.9 TOPS/W at 2 bits, to 2.6 TOPS/W at 3
bits, 1.4 TOPS/W at 4 bits, 0.78 TOPS/W at 5 bits, and 0.24 TOPS/W at 6
bits, so each additional bit costs a factor in efficiency. The symbol
rate of the best design falls from 35 GBaud at 2 bits to 5 GBaud at 6
bits (panel g), but the efficiency is relatively flat over a wide range
of symbol rates. The efficiency of any design is within 20\% of
\(\eta^*\) from 12.5 to 135 GBaud at 2 bits, and from 8 to 55 GBaud at 5
bits. Two effects favour slow channels at high resolution: the sig--sp
SNR of Eq.~\ref{eq:osnr_penalty} grows as \(1/f_B\) through the noise
bandwidth, and the converter power grows as \(2^{B_{conv}} f_B\).

The channel count, in contrast, has almost no effect on efficiency once
\(N\) is above a few channels. In Fig.~\ref{fig:design_space}(f), the
curves rise sharply from low efficiency when \(N=1\) until
\(N \approx 5\)--10 where they remain quite flat. From 2 to 4 bits,
every \(N\) from about 9 up to 64 gives an efficiency within 20\% of
\(\eta^*\), and at 5 and 6 bits it's every \(N\) from 12 to 64.

The four panels in Fig.~\ref{fig:design_space}(b--e) provide a
satisfying explanation for this fact. At a fixed \(N\), the efficiency
rises with the spectral occupancy \(N f_B\), because the fixed power of
each device (laser threshold, modulator bias, detector bias) is shared
over more operations per second. It rises up until it approaches the
spectral limit of the resolution target (Eq.~\ref{eq:wall}), and then
falls steeply close to it. The best design at each \(N\) therefore sits
at a near-constant occupancy just inside this limit. For example, the
optima sit at \(N f_B=0.95\)--1.2 THz (limit 1.35 THz) at 2 bits, and
0.6--1.0 THz (limit 1.06 THz) at 4 bits. Along this curve, the power per
operation depends very little on how the occupancy is split between
\(N\) and \(f_B\), so the curves for different \(N\) almost coincide.
Beyond 4 bits, the effect of the amplifier is more prominent and sets
the occupancy instead. Once the design has a saturated amplifier stage
with gain \(G\) and noise figure \(F\), the signal--spontaneous SNR of
Eq.~\ref{eq:osnr_penalty} limits the occupancy to
\(N f_B \le m^4 P_{sat}/(3 F G h\nu\, 4^{B})\), which is 0.92 THz at 5
bits and 0.23 THz at 6 bits, close to the occupancy of the best designs
(about 0.8 and 0.2 THz). The location of the optimum \(N^*(B)\) within
this plateau is therefore not robust. For a system designer, this result
means that \(N\) can be chosen for throughput, at little cost in
efficiency.

Spatial fan-out on the other hand has a different relationship, because
\(S\) multiplies throughput at fixed spectral occupancy and pays the
same broadcast-loss and saturation penalties as \(N\), but it doesn't
contribute to spectral crosstalk. The optimum fan-out is small and falls
with the target bit resolution from \(S^* = 8\) at 2 bits to 2 from 4
bits upward. At 2 to 4 bits, the efficiency is still relatively flat in
\(S\). Every \(S\) from 2--4 up to 10--16 is within 20\% of \(\eta^*\).
At 5 and 6 bits, the space is only flat up to \(S \approx 4\). From 5
bits, the second amplifier stage sits after the \(S\)-way splitter, so
its input power, and with it the signal--spontaneous SNR, falls as
\(1/S\). Each doubling of \(S\) then costs as much SNR as a doubling of
\(f_B\), so \(S\) no longer increases \(T\) at the same fixed
resolution, while the number of modulators, receivers, and ADCs still
grows with it, reducing the achievable efficiency. At 2 to 4 bits the
most efficient TFLN designs use a single amplifier stage after the first
modulator bank, tuned to about 15 dB of gain, and the TIA noise is
80--90\% of the noise. The receivers and data converters then take the
largest share of the power (39--56\%), and the optical amplifiers take
21--38\% (Tbl.~\ref{tbl:frontier_designs}). From 5 bits up, a second
amplifier stage after the second modulator bank is necessary,
signal--spontaneous noise becomes 80--90\% of the noise, and the
amplifiers take 57--69\% of the power.

\subsection{Resolution--Throughput--Efficiency
Frontier}\label{sec:frontier}

\begin{figure*}
\centering
\pandocbounded{\includegraphics[keepaspectratio,alt={Pareto optimal frontiers. Frontier of resolution versus (a) throughput and (b) compute efficiency for the SiPh and heterogeneous TFLN scenarios using the fully constrained model. The multi-chip TFLN scenario with increased coupling losses is shown for comparison. In (a), the thick lines have no power limit while the thin lines of the same colour impose a realistic power constraint of 1kW on each design. The grey lines show the upper limit to the resolution at N = S = 16, 32 and 64, with no platform-specific constants, link budget, or power limits, for comparison. The same lines can't be plotted for (b) since efficiency can't be evaluated without power consumption. Markers on both plots show contemporary digital accelerators at their stated arithmetic formats (dense, per chip numbers).}]{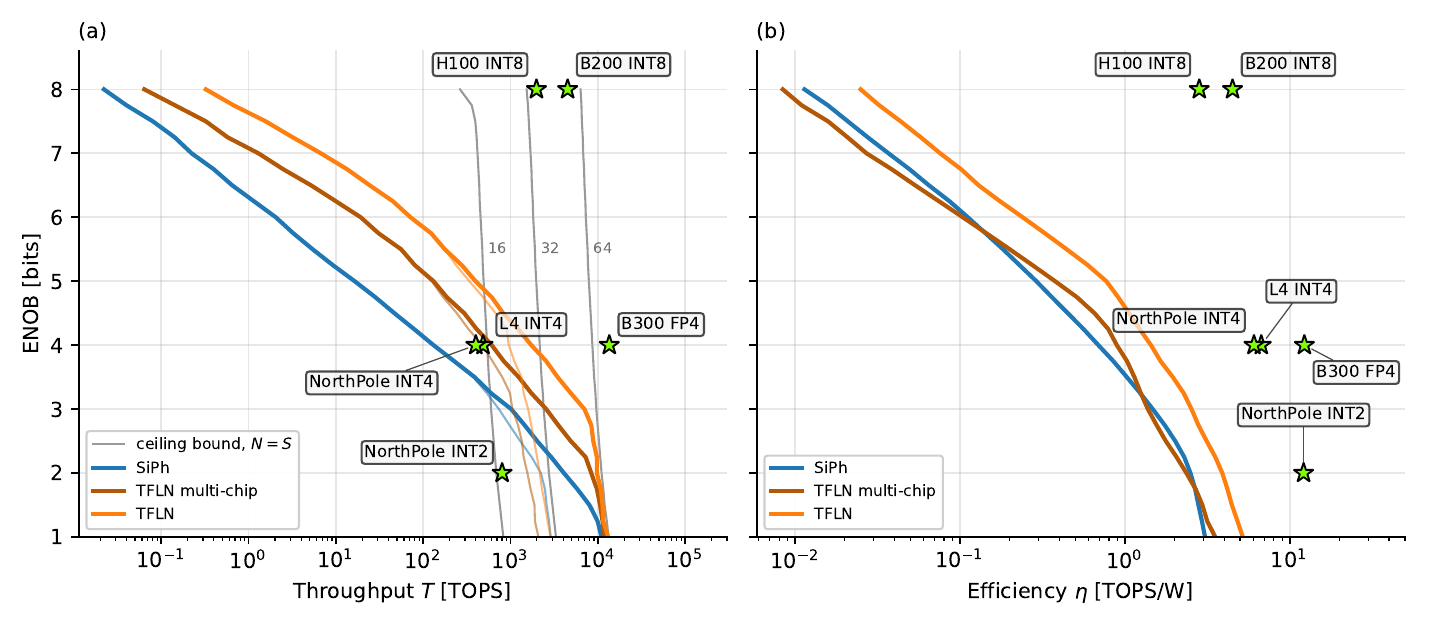}}
\caption{Pareto optimal frontiers. Frontier of resolution versus (a)
throughput and (b) compute efficiency for the SiPh and heterogeneous
TFLN scenarios using the fully constrained model. The multi-chip TFLN
scenario with increased coupling losses is shown for comparison. In (a),
the thick lines have no power limit while the thin lines of the same
colour impose a realistic power constraint of 1kW on each design. The
grey lines show the upper limit to the resolution at \(N = S = 16\), 32
and 64, with no platform-specific constants, link budget, or power
limits, for comparison. The same lines can't be plotted for (b) since
efficiency can't be evaluated without power consumption. Markers on both
plots show contemporary digital accelerators at their stated arithmetic
formats (dense, per chip numbers).}\label{fig:pareto}
\end{figure*}

Fig.~\ref{fig:pareto} combines the above analysis and presents the
optimized frontiers of resolution against throughput and against compute
efficiency under the fully constrained model. For each target
resolution, we search over all designs (\(N\), \(S\), \(f_B\)) which
meet the target (as described in Sec.~\ref{sec:link_budget}), and the
efficiency frontier is the curve representing the most efficient of
these designs, while the throughput frontier is the curve of the designs
with the highest throughput, regardless of power consumption. As a check
on the high power consumption side of the throughput frontier, we also
include a limit on the total power consumption of 1kW, which we plot as
the thin line of the same colour. Each frontier is computed using
amplifier stages with variable gain, integer device counts, optical
power limits for waveguides and amplifier saturation, and data
converters sized for the target resolution. Under the parameters typical
of a heterogeneous TFLN architecture, the efficiency frontier passes
through 3.9, 2.5, 1.4 and 0.77 TOPS/W at 2, 3, 4 and 5 bits. The silicon
scenario, with on-chip amplifiers at 0.41 W per stage and an 8 dB noise
figure, is 36--62\% less efficient. The multi-chip TFLN scenario, with
6.5 dB more of fixed coupling losses and 1 dB more loss per amplifier
stage, gives 2.4, 1.4, 0.89 and 0.38 TOPS/W, which is 38--51\% less
efficient than the heterogeneous baseline.
Tbl.~\ref{tbl:frontier_designs} lists the most efficient designs found
from the frontier search, with power consumption and the maximum
resolution listed for each one.

The first notable result is that amplifier saturation is a binding
constraint in every design, because it limits the power that reaches the
detectors. Up to 4 bits, the most efficient TFLN designs use a single
amplifier stage, and the TIA noise is 80--90\% of their noise. From 5
bits, they need a second stage, signal--spontaneous noise is 80--90\% of
the noise, and the amplifiers take 57--69\% of the power. In the SiPh
designs, with their higher noise figure, signal--spontaneous noise is
the largest noise at every target (67--93\%). The best achievable
compute efficiency therefore depends on the power consumption of the
amplifiers, and more strongly at high resolution. If each TFLN amplifier
stage drew half the power of the measured device, with the same gain,
noise figure and saturation power, the efficiency would rise by a factor
of 1.1 at 2 bits and 1.5 at 6 bits. At twice the power, it would fall by
17--36\%. The second salient point is that spectral crosstalk is a minor
part of the noise of the most efficient designs. The most efficient
2-bit TFLN design runs at \(N f_B = 1.1\) THz, below the 2-bit limit of
1.35 THz, and crosstalk is 8\% of its noise. Crosstalk is 13\% of the
noise at 3 bits, 4\% at 4 bits and 1\% or less above, and the crosstalk
ceiling of the most efficient designs is 1.5 to 3.4 bits above the
target. Third, the throughput frontier reaches GPU-class throughput only
at low resolution and very high power. The TFLN design reaches about
9,700 TOPS at 2 bits using \(N = S = 64\), but it requires about 6.3 kW
to reach that point. Under a 1 kW limit, the TFLN throughput is about
2,200 TOPS at 2 bits, 1,600 TOPS at 3 bits and 950 TOPS at 4 bits. For
comparison, the grey lines of Fig.~\ref{fig:pareto}(a) apply only the
power-independent upper limit on the resolution. At the largest scale of
the search (\(N = S = 64\)), this bound is about 11,000 TOPS at 2 bits,
8,600 TOPS at 4 bits and 6,300 TOPS at 8 bits. This bound is only 1.1
times the TFLN frontier at 2 bits but rises to 5.0 times at 4 bits, so
the designs can only come close to the spectral limit at low resolution.
From about 4 bits, the link and the power set the gap to the digital
accelerators.

\begin{table*}[tbp]
\centering
\caption{Most efficient designs at each target resolution, with the amplifier
stages of Tbl.~\ref{tbl:hardware_params} (1.12 W per stage at full gain
for TFLN, 0.41 W for SiPh). Amplifier saturation is the binding
constraint in every design: 3 dB more saturation power raises the
efficiency by 16--39\% for TFLN and 7--28\% for SiPh. The TFLN source
runs at its 13 dBm cap, but neither a higher source cap nor higher
waveguide caps change the TFLN efficiency by more than 3\% up to 5 bits
(a higher source cap gives 17\% at 6 bits). The multi-chip TFLN designs
behave in the same way; 3 dB more saturation power gives 14--26\% up to
5 bits, the waveguide caps change the efficiency by at most 1\%, and a
higher source cap gives at most 2\% below 5 bits and 13\% at 5 bits. For
SiPh, the 17 dBm waveguide cap also binds at every target (3 dB more
gives 4--23\%). All values are from the sampled fixed-target search of
Sec.~\ref{sec:link_budget}.}\label{tbl:frontier_designs}
{\def\LTcaptype{} 
\begin{tabular}{@{}
  >{\raggedright\arraybackslash}p{(\linewidth - 20\tabcolsep) * \real{0.0938}}
  >{\centering\arraybackslash}p{(\linewidth - 20\tabcolsep) * \real{0.1094}}
  >{\centering\arraybackslash}p{(\linewidth - 20\tabcolsep) * \real{0.0391}}
  >{\centering\arraybackslash}p{(\linewidth - 20\tabcolsep) * \real{0.0391}}
  >{\centering\arraybackslash}p{(\linewidth - 20\tabcolsep) * \real{0.1172}}
  >{\centering\arraybackslash}p{(\linewidth - 20\tabcolsep) * \real{0.0938}}
  >{\centering\arraybackslash}p{(\linewidth - 20\tabcolsep) * \real{0.0703}}
  >{\centering\arraybackslash}p{(\linewidth - 20\tabcolsep) * \real{0.1328}}
  >{\centering\arraybackslash}p{(\linewidth - 20\tabcolsep) * \real{0.1328}}
  >{\raggedright\arraybackslash}p{(\linewidth - 20\tabcolsep) * \real{0.1016}}
  >{\centering\arraybackslash}p{(\linewidth - 20\tabcolsep) * \real{0.0703}}@{}}
\toprule\noalign{}
\begin{minipage}[b]{\linewidth}\raggedright
Platform
\end{minipage} & \begin{minipage}[b]{\linewidth}\centering
Target \(B^*\)
\end{minipage} & \begin{minipage}[b]{\linewidth}\centering
\(N\)
\end{minipage} & \begin{minipage}[b]{\linewidth}\centering
\(S\)
\end{minipage} & \begin{minipage}[b]{\linewidth}\centering
\(f_B\) {[}GBaud{]}
\end{minipage} & \begin{minipage}[b]{\linewidth}\centering
\(T\) {[}TOPS{]}
\end{minipage} & \begin{minipage}[b]{\linewidth}\centering
\(P\) {[}W{]}
\end{minipage} & \begin{minipage}[b]{\linewidth}\centering
\(\eta\) {[}TOPS/W{]}
\end{minipage} & \begin{minipage}[b]{\linewidth}\centering
\(\dfrac{P_{SOA}}{P_{tot}}\)
\end{minipage} & \begin{minipage}[b]{\linewidth}\raggedright
Noise regime
\end{minipage} & \begin{minipage}[b]{\linewidth}\centering
Ceiling
\end{minipage} \\
\midrule\noalign{}
& 2 b & 32 & 8 & 34 & 551 & 141 & 3.90 & 21\% & TIA & 3.8 b \\
& 3 b & 30 & 4 & 34 & 242 & 95 & 2.54 & 29\% & TIA & 4.5 b \\
TFLN & 4 b & 35 & 2 & 25 & 122 & 85 & 1.44 & 38\% & TIA & 6.3 b \\
& 5 b & 48 & 2 & 16 & 148 & 191 & 0.77 & 57\% & sig--sp & 8.0 b \\
& 6 b & 45 & 2 & 4.9 & 40 & 165 & 0.24 & 69\% & sig--sp & 9.4 b \\
& & & & & & & & & & \\
& 2 b & 14 & 4 & 71 & 111 & 47 & 2.36 & 32\% & TIA & 4.7 b \\
TFLN & 3 b & 15 & 2 & 61 & 55 & 40 & 1.38 & 40\% & TIA & 5.7 b \\
multi-chip & 4 b & 42 & 2 & 22 & 152 & 170 & 0.89 & 64\% & sig--sp, TIA
& 5.9 b \\
& 5 b & 42 & 1 & 14 & 49 & 129 & 0.38 & 68\% & sig--sp & 8.6 b \\
& & & & & & & & & & \\
& 2 b & 63 & 8 & 17 & 1078 & 431 & 2.50 & 30\% & sig--sp & 3.9 b \\
& 3 b & 39 & 4 & 17 & 206 & 140 & 1.48 & 35\% & sig--sp & 8.5 b \\
SiPh & 4 b & 21 & 4 & 11 & 39 & 57 & 0.69 & 40\% & sig--sp & 8.8 b \\
& 5 b & 12 & 2 & 8.9 & 5.1 & 18 & 0.29 & 38\% & sig--sp & 8.9 b \\
& 6 b & 7 & 2 & 5.2 & 1.0 & 9.2 & 0.11 & 39\% & sig--sp & 9.3 b \\
\bottomrule
\end{tabular}
}
\end{table*}

The markers in Fig.~\ref{fig:pareto} show how the PTC compares to
several contemporary digital accelerators. We list the commercial values
for the NVIDIA H100 (INT8, 1979 TOPS, 700 W) {[}31{]}, B200 (INT8, 4500
TOPS, 1000 W) {[}32{]}, B300 (FP4, 13,500 TOPS, about 1100 W) {[}33{]},
and L4 (INT4, 484 TOPS, 72 W) {[}34{]}, as well as the IBM NorthPole
research chip (INT4, about 400 TOPS, and INT2, about 800 TOPS, at about
66 W measured board power) {[}35{]}, which aid as a digital reference
point for low-precision fixed-point formats. The data points are not
exactly comparable to the computed ENOB of the PTC, but the points
illustrate the gap between current PTC capabilities and the
state-of-the-art digital SIMD paradigm. The ENOB of Eq.~\ref{eq:enob} is
the resolution of an ideal quantizer on the output, which is not the
same as digital INT-\(b\) arithmetic, where the operands are quantized,
and the products are accumulated exactly. For the same output error,
\(b\)-bit operand quantization corresponds to an ENOB that is 0.29 bits
higher for uniform operands, and about 0.8--1.1 bits higher for sparse
or Gaussian operands (Sec.~\ref{sec:methods}), so the comparison is
approximate. Furthermore, the floating-point format FP4 is not four
uniform bits, so it isn't directly comparable to the fixed-point
representation used in analog PTCs, but is included for perspective for
state of the art throughput and efficiency. In terms of power
consumption, the model developed here attempts to undertake an accurate
accounting of all the components of the compute core, such as sources,
modulators, drivers, amplifiers, receivers, converters, and thermal
stabilization, but it excludes peripherals beyond the analog domain such
as memory, host interface, and system cooling. The figures for the
digital accelerators are published package-level thermal-design power,
so our efficiency axis is optimistic in scope even where its parameters
are conservative.

Under these constraints, we see that the efficiency of the
hypermultiplexed photonic tensor core reaches the 2.8--4.5 TOPS/W of the
INT8 accelerators only at 2 bits of effective resolution (3.9 TOPS/W for
TFLN, while SiPh reaches 2.5 TOPS/W), which is 6 bits below their
format, although the throughput is higher than some digital designs. At
4 bits the TFLN frontier (1.4 TOPS/W) is 4.2 to 4.7 times below the INT4
accelerators (6.1--6.7 TOPS/W) and a factor of 8.5 below the B300 at FP4
(12.3 TOPS/W). At 2 bits, NorthPole at INT2 (about 12 TOPS/W) is about
3.1 times more efficient than the best photonic design. The highest
resolution in our search is 8 bits, which both PTC platforms reach only
with very small designs, at 0.025 TOPS/W for TFLN (\(N = S = 3\)), and
0.011 TOPS/W for SiPh (\(N = 2\), \(S = 1\)).

\section{Discussion}\label{sec:discussion}

The results of this analysis reveal several interacting constraints on
the design of PTCs, three of which are physical limits of the
architecture. The \(10\log_{10}(NS)\) broadcast loss is a fundamental
penalty to the passive copying of an analog signal which degrades the
effective bit resolution of the computation. Relatedly, the limits to
the aggregate power in a waveguide or in an amplifier push the gain
further down the optical link as \(N\) and \(S\) grow, and this degrades
the noise performance of the optical amplifiers, regardless of their
technology. Lastly, even allowing the possibility for mitigating these
constraints with higher saturation limits and lower noise figures, this
model identifies the power-independent ceilings on the resolution set by
the source RIN and the spectral crosstalk due to the shape of the
modulation PSD and demultiplexer transmission response. A fourth
fundamental constraint, which arises due to the encoding itself rather
than the architecture, is the resolution penalty due to the encoding of
the result as a product of intensities. The result of the computation,
the dot product, is only the useful fraction of the signal given by
\(\sigma_d^2 m^4\) (\(-9.5\) dB at \(m = 1\)), while the dominant noise
components beat against the entire signal. This penalty is shared by any
architecture which employs a single carrier onto which both operands are
modulated. Beyond these fundamental limits, we identify that practical
constraints of current technology platforms---coupling and insertion
loss, data converter figures of merit, receiver power consumption, and
amplifier saturation power and noise figure---pose significant
limitations to the achievable resolution and efficiency, reducing it
below the constraints of the architecture and encoding. In this
architecture, the constraints of the spectrum and the quality of the
laser source set the upper limit to the resolution, but with practical
components available on current technology platforms, the amplified link
constraints set a more restricted upper limit. Together with the power
consumption of the devices, the details of the technology platforms used
for the photonic link constrain the optimal values for the fan-out
\(S^*(B)\), the symbol rate \(f_B^*(B)\), and the operating point on
Eq.~\ref{eq:snr_ceiling}. Above about ten channels, the efficiency
hardly depends on \(N\), and the spectral limit to the channel density
is at most a minor part of the noise of the most efficient designs,
largest at 2 and 3 bits of resolution.

Other similar architectures avoid the spectral limit through different
encodings. In coherent photoelectric multiplication, inputs and weights
are both encoded as field amplitudes in time, their product is achieved
with a homodyne detector, and charge accumulation performs the sum over
symbols {[}36{]}; this scheme only makes use of spatial parallelism with
a single wavelength, so the precision limit is given by the shot noise
of detected photons, rather than crosstalk or amplifier noise.
Brückerhoff-Plückelmann et al.~proposed instead to sum 16 wavelengths
and 1000 symbols on one slow detector, and projected 110 TMAC/s from 100
GBaud modulators on a 100 GHz grid {[}17{]}; in this case the beat term
between channels on one detector falls outside of its bandwidth, so the
spectral limit only applies between the groups of channels that the
demultiplexer separates. Neither of these works reports a precision at
its projected scale, so we cannot place them in
Fig.~\ref{fig:wall_master}(a).

The classification of noise terms into terms independent of, linear in,
and quadratic in the photocurrent, as in Eq.~\ref{eq:master_snr} is a
standard method in analog photonic links for grouping noise
contributions {[}27, 28{]}. There have also been previous scaling
analyses for photonic accelerators which link SNR to precision {[}26,
37{]}; the model presented here adds three main things beyond these
previous works. We characterize the content of the upper resolution
limit for wavelength-multiplexed computing architectures, specifically
hypermultiplexed PTCs, given by the cross-talk coefficient related to
the data statistics of neighbouring channels and the resolution penalty
imposed by encoding the product using cascaded modulators. We also
define a limiting relationship between the packing of spectral channels
and the resolution of the computation. Last, we undertake our analysis
in two stages, first presenting a universal model for the resolution in
the receiver of a PTC, and subsequently applying it to an exact worked
example of the link budget of an amplified photonic link where we
optimize amplifier placement and source power using realistic parameters
from current technology platforms.

The reported precision of the published systems shown as markers in
Fig.~\ref{fig:wall_master}(a) sit well below the predicted ceiling ENOB
for their size. In the largest scale demonstration of this architecture
to date, Pappas et al.~report a measured ENOB of 1.7 bits (NEB 3.5)
against a 2.4 bit ceiling at 32 GBaud. Their lines are spaced by 100
GHz, which is closer than the \(\Delta\nu_{tot}/N = 250\) GHz that the
map of Fig.~\ref{fig:wall_master}(a) assumes at \(N = 16\) which may be
the reason the ceiling for their system sits lower lower than the
analysis suggests given their scale. At 20 GBaud, the same spacing gives
a ceiling of 7.5 bits, far above their measured ENOB of 2.5 bits (NEB
4.3), so cross-talk clearly does not limit their precision at this rate.
To estimate which factors likely constrain their implementation, we
applied the link model of Sec.~\ref{sec:link_budget} to their testbed,
using parameters typical of a discrete setup (EDFA-class amplifiers with
appropriate coupling loss), rather than the integrated scenarios
considered in this work, with \(N = S = 16\), 100 GHz spacing. The
highest ENOB that the model allows for this scenario is 4.7 bits at 20
GBaud, limited by signal--spontaneous beat noise, and 2.3 bits at 32
GBaud, limited by cross-talk. Both published measurements sit 0.6--2.2
bits below these predictions, which is expected for a model that leaves
out the modulator nonlinearity, the noise of the converters and lab
instruments, intersymbol interference and the in-band cross-talk of the
router.

Compared with digital accelerators, we resist both triumphal and
defeatist readings of our results. Under the stated limitations of the
comparison of the current model to digital accelerators---the IEEE ENOB
convention and neglecting power consumption beyond the compute
system---the PTC designs considered can match the efficiency of INT8
accelerators (2.8--4.5 TOPS/W) only at an effective 2 bit computation,
and at 4 bits, the TFLN frontier lies below the most efficient INT4
accelerators by a factor of 4.2 to 4.6 (8.4 for the B300 using the FP4
format). The most efficient designs of Tbl.~\ref{tbl:frontier_designs}
clearly show where improvements must come from in order to close the gap
compared with digital accelerators. Optical amplifier saturation limits
the power that reaches the detectors in every design. Up to 4 bits, the
receivers and converters take the largest share of the power of the TFLN
designs (40--58\%) and the TIA noise sets the resolution; from 5 bits
onward, the amplifiers take most of the power (57--69\%) and
signal--spontaneous noise sets the resolution. For PTC architectures
that fall under our model, the largest improvements will therefore come
from lower coupling and insertion losses and from optical amplifiers
with higher saturation power per Watt of power consumption, and a
reduction in the noise figure. Removing 6.5 dB of fixed loss and 1 dB
per amplifier stage (the difference betwee our heteregeneous and
``multi-chip'' scenarios) raises the efficiency by 1.6 to 2.1 times.
Halving the power of the amplifiers alone gives 1.5 times the efficiency
at 6 bits. At low target resolutions, lower receiver power, reduced TIA
noise, and lower data converter power are the next levers, while a
better source RIN would have little effect at all points of the design
space. Techniques to limit the spectral crosstalk, such as pulse
shaping, or demultiplexers with steeper roll-off, help mainly at 2 and 3
bits, where crosstalk is the largest share of the noise.

Sec.~\ref{sec:resolution_results} covers the costs and benefits of pulse
shaping symbols in this type of PTC. Pulse shaping moves the spectral
limit of Eq.~\ref{eq:wall}, but the magnitude of its effect is set by
whichever part of the channel has the longest tail between the signal
PSD (including its broadening due to being encoding a product), the
demultiplexer transmission profile, or the receiver's electrical
response. Shaping is also not free in an analog PTC, since the overshoot
of band-limited waveforms costs modulation depth (which the product
encoding pays at fourth order), and every one of the \((N + SN)\)
transmitters needs its own digital filter and oversampled DAC. Because
crosstalk is at most a minor part of the noise of the most efficient
designs, we leave the optimization of the shaping strength for future
work, and note that optical telecommunications sits in the opposite
corner of this design space, where a few valuable channels can easily
carry the cost of heavy shaping.

The principal limitations of this analysis are the deferred effects
listed in Sec.~\ref{sec:methods}. Nonlinearity in the modulator and its
driver will add distortion components which contribute to the noise
budget at large modulation indices when there is no, or imperfect,
pre-distortion applied. Likely the most important deferred cross-talk
effect is the in-band crosstalk of the wavelength router. A fraction
\(\varepsilon\) of the victim's own wavelength leaks from each of the
other \(N - 1\) input ports, and can't be filtered out because it's
directly in the band the receiver is tuned to. Because all ports share
the same source lines, the leak is coherent with the victim and the
error scales as \(\sqrt{\varepsilon}\), not \(\varepsilon\) as the sum
of intensities from {[}20{]} suggests. Most of this error is static
though, since it would be comprised of an offset and a fixed mixing of
all the undesired weight rows, so in principle a calibration and
pre-compensation of the programmed weights could remove it.

Another limitation in this analysis is the crosstalk coefficient
\(k_{xt}\); a numerical simulation of the field with bounded operands
supports the first-order expression up to a channel spacing of
\(5 f_B\), and we keep a margin \(\kappa = 1.5\), but the true spectrum
of the intensity modulation, and its effect on the far tail that sets
the crosstalk at wide channel spacing, depends on the pulse shaping and
the driver response, which have not been measured. The other deferred
effects of Sec.~\ref{sec:model_scope} can move the numerical values or
change the location of the optima found in this analysis, but we believe
that most of these deferred effects will move the numbers less than the
technology parameters do.

The modelling framework put forward in this work is the most widely
applicable contribution. The receiver model of Eq.~\ref{eq:master_snr}
does not depend on the architecture: any wavelength-multiplexed IM/DD
scheme with the same signal and noise assumptions delivers a
\((P_{det}, S_{ASE}, \chi, RIN)\) tuple to its detectors and lands on
the same universal curve. For any given platform, only the specifics of
the technology platform and the link need to be defined. There are a
number of natural extensions of this work, which could each be readily
calculated given the present framework. Expansions of the optical
spectrum beyond the C-band (about 4 THz wide) are one such extension;
moving to the O-band (17 THz) would represent a direct multiplication of
the spectral limit in Eq.~\ref{eq:wall} by a factor of just over 4.
Detailed studies of specific architectures with measured results could
further validate the model. Mixed-technology gains (for example, using
cheaper and lower power amplifiers further down the link versus higher
power and lower noise links further upstream) could be tested using this
framework, and together with a gain model in which the noise figure, the
saturation power and a transparency current change with the bias of each
stage, would constitute the biggest improvement to the amplifier
performance under current technology.

\section{Methods}\label{sec:methods}

Beyond the raw throughput of the PTC, we are interested in its compute
efficiency, and therefore its power consumption, as well as the
effective number of bits (ENOB) delivered at the output of the
computation. The power consumption of the active components, the various
sources of noise they introduce, and the loss experienced by all the
signals in all passive components, are all intricately linked to the
same scaling parameters that govern the throughput (\(N\), \(S\), \(K\),
and \(f_B\)).

Our model of the ENOB and compute efficiency of the hypermultiplexed PTC
is organized in two parts. The first part
(Sec.~\ref{sec:receiver_model}, Sec.~\ref{sec:spectral_model}) needs no
knowledge of the architecture or of the technology-dependent power
figures. No matter the details of the optical front-end, each
photodetector receives two quantities set by the details of the link,
the average detected power \(P_{det}\) and the accumulated
amplified-spontaneous-emission (ASE) spectral density \(S_{ASE}\), and
two properties of the signal itself, the relative intensity noise of the
source and the inter-channel crosstalk ratio \(\chi\). The
signal-to-noise ratio, and hence the ENOB, is a closed-form function of
these quantities alone. The second part (Sec.~\ref{sec:link_budget},
Sec.~\ref{sec:power_model}) is specific to an architecture using a
defined set of technologies (power consumption values, loss metrics).
The link budget relates the scaling parameters and amplification choices
to \((P_{det}, S_{ASE})\), and the models of individual components are
then used to accumulate a total power consumption figure. To then
explore the design space, these two parts can be combined with an
explicit goal for each design. For any given value of the scaling
parameters, the model leaves the source power and the amplifier gain
unconstrained, but both values affect the noise and the power
consumption of the whole system. At each design point (combination of
\(N\), \(S\), \(f_B\)), we resolve this by choosing the source power,
the amplifier placement, and amplifier gain value to minimize the total
electrical power consumption, subject to a target ENOB
(Sec.~\ref{sec:link_budget}). All efficiency results in
Sec.~\ref{sec:results} use this policy. The resolution ceiling and the
link-limited resolution report the maximum achievable ENOB instead. The
numerical values of all parameters are listed in
Sec.~\ref{sec:instantiation} (Tbl.~\ref{tbl:hardware_params}).

\subsection{Signal and Noise at the Detector}\label{sec:receiver_model}

\subsubsection{Signal Model}\label{signal-model}

Each optical channel passes through two MZMs in series, both biased at
quadrature. The first encodes a weight \(w(t)\) and the second encodes
an input \(x(t)\), with \(w, x \in [-1, 1]\) updated at the symbol rate
\(f_B\). Linearizing each quadrature-biased transfer function, the
detected optical power on one (wavelength, output-port) pair is

\begin{equation}\protect\phantomsection\label{eq:cascaded_mzm}{
\begin{aligned}
p(t) &= P_{det}\left[1 + m\,w(t)\right]\left[1 + m\,x(t)\right] \\&= P_{det}\left[1 + m w + m x + m^2 w x\right],
\end{aligned}
}\end{equation}

where \(P_{det}\) is the average detected power (the quadrature bias
losses are carried in the link budget, Sec.~\ref{sec:link_budget}) and
\(m\) is the intensity-modulation index. We take \(m = 1\), the full
intensity range of the modulator. A linear intensity response over this
range needs a predistorted drive with a peak phase excursion of
\(\arcsin m\), so \(m = 1\) needs the full swing \(V_{pp} = V_\pi\),
which has a power consumption cost covered in
Sec.~\ref{sec:power_model}. The predistortion itself is a fixed
nonlinear map applied before the DACs for which we have not evaluated
the costs, but in principle could be done at very little energy cost
with a lookup table in the transmitter. A smaller index---for example,
the operating point \(V_{pp} \approx 0.8 V_\pi\) without predistortion
used in {[}23{]}---keeps the drive near the linear part of the transfer
function, but the product encoding incurs a penalty of \(m^4\) in signal
power (3.9 dB at \(m = 0.8\)).

The useful computational signal is the product term, with photocurrent
\(i_{sig}(t) = m^2 I_{ph}\, w(t)x(t)\), which has variance,
\begin{equation}\protect\phantomsection\label{eq:signal_power}{
\langle i_{sig}^2 \rangle = \sigma_d^2\, m^4 I_{ph}^2, \qquad \sigma_d^2 \equiv \langle w^2 \rangle \langle x^2 \rangle,
}\end{equation} where \(I_{ph} = \mathcal{R} P_{det}\) is the average
photocurrent and \(\mathcal{R}\) is the responsivity.
Eq.~\ref{eq:signal_power} makes the dependence on the data statistics
explicit; for uniform independent inputs on \([-1,1]\),
\(\sigma_d^2 = 1/9\), which is the value used throughout the analysis.
Other operand statistics, such as Gaussian, or the sparse, long-tailed
distributions typical of activations in deep neural networks, have
smaller \(\sigma_d^2\). We have not computed the resulting change in SNR
and range here, so the ENOB values hold only in the case of uniform
operands.

In Eq.~\ref{eq:cascaded_mzm}, the constant term and the linear terms
\(m w(t)\) and \(m x(t)\) are undesired components that must be removed
to extract the result of the computation. The constant term is easily
removed with AC coupling, but the linear terms are in the same frequency
band as the product, so they cannot be removed in the same way. After
removal of the constant term and accumulation over \(L\) symbols, the
remaining terms hold
\(m \sum_l w_l + m \sum_l x_l + m^2 \sum_l w_l x_l\). The term
\(\sum w_l\) is usually known in advance, since the weights are
programmed, so it can be removed deterministically using an analog bias
current in the receiver, whereas \(\sum x_l\) must be forced to zero
using a different technique. This can be achieved either with a balanced
encoding which forces the term to have a zero mean, as performed in
{[}23{]}, or by using a modification of the architecture to use a
constant signal on a different wavelength injected into the second input
of each MZM in the second bank, along with a balanced photodetector on
the output, as performed in {[}22{]}. The linear terms are larger than
the product term by about \(1/m\), so removing them before digitization
is necessary to avoid requiring extra bits in the ADC. If the
subtraction of the linear terms were performed digitally, the ADC would
have to cover a range \(1 + 2/m \ge 3\) times larger, which costs
\(\ge \log_2 3 \approx 1.6\) extra bits and 3 times the power under the
model used in Sec.~\ref{sec:power_model}.

\subsubsection{Noise Model}\label{noise-model}

The receiver integrates the photocurrent over \(L\) symbol periods
\(1/f_B\), samples the result, and then resets for the following cycle.
The sampled output is the sum of \(L\) products, and the signal
variance, the noise variance, and the full-scale range all grow with
\(L\). The SNR of one MAC operation therefore does not depend on \(L\);
it is just the SNR of one symbol, evaluated at the noise bandwidth of a
one-symbol integrator, which has power response
\(\mathrm{sinc}^2(f/f_B)\),
\begin{equation}\protect\phantomsection\label{eq:noise_bandwidth}{
\Delta f = \int_0^\infty \mathrm{sinc}^2(f/f_B)\,df = \frac{f_B}{2}.
}\end{equation} All noise spectral densities listed in this section are
one-sided and they all are defined with respect to this same noise
bandwidth \(\Delta f\).

Every SNR and ENOB in this analysis uses this per MAC convention
referred to the operand range. The resolution of the ADC on the other
hand must be referred to the range of the full scale output, which is
\(\tfrac{1}{2}\log_2 L\) bits higher, because the range grows as \(L\)
and the noise as \(\sqrt{L}\) (Sec.~\ref{sec:power_model}). The
converters are sized one bit above the target, so we leave their
quantization noise out of the budget.

The total noise power at the receiver is the sum of independent
contributions from all the effects considered,
\begin{widemath}\begin{equation}\protect\phantomsection\label{eq:noise_sum}{
\langle i_n^2 \rangle = \langle i_{TIA}^2 \rangle + \langle i_{shot}^2 \rangle + \langle i_{RIN}^2 \rangle + \langle i_{sig\text{-}sp}^2 \rangle + \langle i_{sp\text{-}sp}^2 \rangle + \langle i_{xt}^2 \rangle,
}\end{equation}\end{widemath}\noindent  each of which is defined in its
corresponding section below.

\emph{Receiver electronics.} TIAs are typically characterized by an
input-referred noise current density \(\bar{\imath}_{TIA}\), which is a
white noise density giving
\(\langle i_{TIA}^2 \rangle = \bar{\imath}_{TIA}^2 \Delta f\). This
figure already includes the Johnson noise of the input stage, so no
separate thermal term is added.

\emph{Shot noise.} The photocurrent variance due to the discrete nature
of charge carriers is
\(\langle i_{shot}^2 \rangle = 2 q I_{ph} \Delta f\), with \(q\) the
electron charge.

\emph{Relative intensity noise.} Noise from a laser source is typically
characterized by a RIN value in dB, which is a relative intensity
fluctuation, leading to a photocurrent variance
\(\langle i_{RIN}^2 \rangle = r\, I_{ph}^2 \Delta f\) with
\(r = 10^{RIN/10}\). RIN is a fixed fraction of the source intensity, so
it is unchanged in relative terms through linear loss and gain terms.
Comb sources may exhibit markedly different per-line RIN than DFB
arrays, but this is not modelled in this work.

\emph{ASE beat noise.} The amplified spontaneous emission in optical
amplifiers gives rise to a single-polarization ASE density \(S_{ASE}\).
The square-law detection of the accumulated ASE field and the signal
together produces beat terms for the signal--spontaneous and
spontaneous--spontaneous combinations {[}38{]},
\begin{equation}\protect\phantomsection\label{eq:beat_noise}{
\begin{aligned}
\langle i_{sig\text{-}sp}^2 \rangle &= 4 \mathcal{R}^2 P_{det} S_{ASE} \Delta f = 4 \mathcal{R} S_{ASE} \Delta f\, I_{ph}, \\\qquad \langle i_{sp\text{-}sp}^2 \rangle &= 2\mathcal{R}^2 S_{ASE}^2 \Delta\nu_o \Delta f,
\end{aligned}
}\end{equation} where \(\Delta\nu_o\) is the demultiplexer's optical
bandwidth (Sec.~\ref{sec:spectral_model}).

\emph{Interchannel crosstalk.} We consider the spectral crosstalk from
neighbouring channels to contribute towards an effective noise variance,
\begin{equation}\protect\phantomsection\label{eq:xtalk_noise}{
\langle i_{xt}^2 \rangle = k_{xt}\, \chi\, I_{ph}^2.
}\end{equation} where \(\chi\) is a leakage factor which describes how
much of the neighbouring channels' spectrum is passed through the victim
signal's demultiplexer and receiver (described in Eq.~\ref{eq:chi}), and
\(k_{xt}\) is a dimensionless coefficient representing the magnitude of
the beat between the victim carrier and the neighbours' modulation
sidebands. For our parameters, \(k_{xt} \approx 0.39\)
(Sec.~\ref{sec:spectral_model}), derived in Appendix
\ref{sec:app_xtalk}.

\subsubsection{Upper Limits to SNR}\label{upper-limits-to-snr}

Grouping the noise terms in Eq.~\ref{eq:noise_sum} by their power in
\(I_{ph}\), together with the signal power of Eq.~\ref{eq:signal_power},
gives Eq.~\ref{eq:master_snr} with the coefficients,
\begin{equation}\protect\phantomsection\label{eq:coefficients}{
\begin{gathered}
A_0 = \left(\bar{\imath}_{TIA}^2 + 2\mathcal{R}^2 S_{ASE}^2 \Delta\nu_o\right) \Delta f, \\
A_1 = \left(2q + 4\mathcal{R} S_{ASE}\right) \Delta f, \\ 
A_2 = r \Delta f + k_{xt}\chi,
\end{gathered}
}\end{equation} where the normalized form with the characteristic
current \(I^* = A_0/A_1\) gives the universal form of
Eq.~\ref{eq:snr_ceiling}. The upper limit to the SNR is given by
\(\mathrm{SNR}_\infty\), which depends only on the term \(A_2\) since
these noise sources are multiplicative ratios on the signal current
itself. \(\mathrm{SNR}_\infty\) depends on \(N f_B\) through \(\chi\),
on \(f_B\) through the RIN, and on the source and filtering shapes which
give rise to \(k_{xt}\).

The classification of the coefficients into constant, linear, and
ceiling terms is defined with \(P_{det}\) and \(S_{ASE}\) treated as two
separate inputs although they are not entirely independent. Turning up
the gain of an amplifier raises \(S_{ASE}\) together with \(P_{det}\),
so sig--sp noise grows like the signal squared and saturates the SNR
just as a ceiling term would, whereas turning up the laser power on its
own leaves \(S_{ASE}\) unchanged and sig--sp grows linearly.

\subsubsection{Effective Number of Bits}\label{effective-number-of-bits}

We use the ENOB definition of IEEE Std 1241 {[}29{]}, which is the
resolution \(B\) of an ideal quantizer, spanning the full-scale range of
the signal \(I_\text{FS}\) with RMS error equal to the RMS noise
\(\sigma_n = \sqrt{\langle i_n^2\rangle}\), so \begin{equation}{
2^B = \frac{I_\text{FS}}{\sqrt{12}\,\sigma_n}.
}\end{equation} An ideal \(B\)-bit quantizer has step size
\(\Delta = I_\text{FS}/2^B\) and associated quantization error variance
\(\Delta^2/12\). The SNR of a signal with variance \(\sigma_{sig}^2\) is
then given by,
\begin{equation}\protect\phantomsection\label{eq:enob_general}{
\mathrm{SNR}_{dB} = 6.02\,B + 10\log_{10}\!\big(3\,\hat{\sigma}^2\big),
}\end{equation} where
\(\hat{\sigma}^2 \equiv \sigma_{sig}^2 / (I_\text{FS}/2)^2\). The
\(+1.76\) dB offset often seen in this equation {[}39{]} is the special
case of a sinusoid which occupies the full range
(\(\hat{\sigma}^2 = 1/2\)). In this work, the signal is the product
\(wx\), which, for independent uniform operands, it spans \([-1,1]\)
with \(\hat{\sigma}^2 = \sigma_d^2 = 1/9\), which translates into the
offset \(10\log_{10}(1/3) = -4.77\) dB (Eq.~\ref{eq:enob}). The
correction for the variance of the signal is important because it
compensates for the definition of the SNR which also depends on the data
statistics, and makes \(B\) invariant with respect to it. For a given
level of noise compared to the full scale range, the ENOB is the same
for any data.

The most common definition of bit resolution seen in the analog
computing literature is the noise-equivalent bit resolution
\(NEB = \log_2(I_\text{FS}/\sigma_n)\) {[}15, 30{]}, which simply counts
the number of bins one \(\sigma_n\) wide across the range. We avoid this
definition because it overstates the precision of a computation compared
to an equivalent digital processor. For a quantizer with step size equal
to the noise standard deviation \(\sigma_n\), values would be mis-read
about 60\% of the time.

The exact digital-equivalent resolution \(b\) is the bit width of the
operands in a digital MAC whose error is equal to \(\sigma_n\). Since
the true operands round to the nearest digital value, they also
experience a quantization error \(e_{x,w}\), and the product has an
error \(w e_x + x e_w\). With some simplification for the ranges of the
products and the operands, we have
\begin{equation}\protect\phantomsection\label{eq:bdig}{
b = B + \tfrac{1}{2}\log_2\big(\langle \tilde w^2\rangle + \langle \tilde x^2\rangle\big),
}\end{equation} where \(\langle \tilde x^2\rangle\) and
\(\langle \tilde w^2\rangle\) are the mean squares of the operands. This
represents a data-dependent correction to the ENOB around equal to
\(-0.29\) bits for uniform operands or \(-0.79\) bits with sparse
operands for example {[}40, 41{]}. We prefer ENOB here because it
doesn't have the data-dependence of \(b\), but this conversion is
important to understand when comparing with digital accelerators.

\subsection{Spectral Packing and Crosstalk}\label{sec:spectral_model}

A channel with intensity modulation at a symbol rate \(f_B\) occupies a
proportional optical bandwidth, so the number of wavelength channels
\(N\) is coupled to total optical bandwidth by
\(\Delta\nu_{ch} = \Delta\nu_{tot}/N\), where \(\Delta\nu_{tot}\) is the
total width of the optical band, e.g.~4 THz in the C-band. Regardless of
\(f_B\), there is a minimum channel spacing imposed by the AWGR, so that
\(\Delta\nu_{ch} \ge \Delta\nu_{ch,min}\).

There are a number of related optical bandwidths to consider, so we
adopt some simple conventions to relate all relevant bandwidths to
\(f_B\). We take the PSD of the modulated signals to have a Gaussian
profile with \(1/e\) half-width \(\Delta\nu_{sig} = a f_B\), where we
set \(a=1\). The demultiplexer is also modelled as a Gaussian power
transmission with width
\(\Delta\nu_o = \max(a_o\Delta\nu_{sig},\, \Delta\nu_{o,min})\), where
\(a_o=2\) is set so that the filter is wide enough to pass the signal
without distortion, and \(\Delta\nu_{o,min} = 20\) GHz is the minimum
feasible bandwidth of a multiplexer channel. After the photodetector, we
take the TIA filter to also have a Gaussian profile with bandwidth
\(\Delta f_{el} = f_B\). Since all the various filters in the signal
path all have a Gaussian shape, we can define a combined receiver
electro-optic filter width, such that
\(\Delta\nu_c^{-2} = \Delta\nu_o^{-2} + \Delta f_{el}^{-2}\), and
\(\Delta\nu_c = 0.89 f_B\), and we are able to algebraically manipulate
the transmission functions to get closed form solutions. Our framework
allows for arbitrary filter shapes and signal PSDs, but for other shapes
the overlap integrals of the derivation in Appendix \ref{sec:app_xtalk}
must be evaluated numerically. Since each filter and PSD is defined with
width related to \(f_B\), the crosstalk (above 10 GBaud, where
\(\Delta\nu_{o,min}\) is no longer a constraint) depends only on
\(\Delta\nu_{ch}/f_B=\Delta \nu_{tot}/(N f_B)\), so the spectral limit
is a limit on the spectral occupancy \(N f_B\). The Gaussian TIA
response with \(\Delta f_{el} = f_B\) has a noise bandwidth of
\(0.89 f_B\), wider than the \(f_B/2\) of Eq.~\ref{eq:noise_bandwidth};
using the \(\mathrm{sinc}^2\) response of that section instead, the
spectral occupancy limit increases slightly below 4 bits but falls by up
to 13\% above 4 bits (0.69 against 0.79 THz at 8 bits), because the
\(\mathrm{sinc}^2\) tail decays only as \(f^{-2}\). We can expect a real
receiver to fall somewhere between these two responses.

The full derivation of the crosstalk model is in Appendix
\ref{sec:app_xtalk}, and yields two dominant spectral crosstalk terms:
the ``beat term'', which comes from the beat of the neighbouring
channels' sidebands with the victim channel's carrier passing through
the demultiplexer and the receiver, and a ``leaked intensity term'',
which comes from the beat between the neighbours' leaked carriers and
their own sidebands. The strength of the beat term is set by the
transmission of the neighbour signal's PSD at an offset
\(\Delta \nu_{ch}\) along with the width of the combined
demultiplexer--receiver filter, \(\Delta \nu_c\). The leaked intensity
term is set mainly by the transmission of the tails of the demultiplexer
at an offset \(\Delta \nu_{ch}\). For \(a_o=2\), the leaked intensity
term is 2.6 to 12 times higher than the beat term for channel spacings
from \(\Delta \nu_{ch}=2\) to 5 \(f_B\). The beat term dominates only
for Gaussian demultiplexers narrower than about \(1.8 f_B\); a more
realistic demultiplexer with a flat pass-band and steeper roll-off, with
an isolation factor of \(\varepsilon\) would have leaked intensity only
on the order of \(\sigma_\mu^2 \varepsilon^2\), which is negligible.
Including all of these terms, the noise variance contributed by the
spectral crosstalk is given by Eq.~\ref{eq:xtalk_noise}, with
\(k_{xt} = \tfrac{1}{2}\sigma_\mu^2 g'\kappa\) and
\(g' = \Delta\nu_c/\sqrt{\Delta\nu_c^2 + \Delta\nu_{sig}^2}\), where the
coefficients are \(\sigma_\mu^2 = 7/9\), \(g' = 2/3\), and
\(\kappa = 1.5\), so \(k_{xt} \approx 0.39\) (see further details in
Appendix \ref{sec:app_xtalk}). The leakage factor \(\chi\), which
combines all of the relevant filter widths and channel spacing, is
defined by Eq.~\ref{eq:chi} in Appendix \ref{sec:app_xtalk}, but can be
simplified by considering only the beat term of each of the two
neighbours, giving,
\begin{equation}\protect\phantomsection\label{eq:chi_beat}{
\chi \approx 2 \exp\!\left(-\frac{\Delta\nu_{ch}^2}{\Delta\nu_{sig}^2 + \Delta\nu_c^2}\right).
}\end{equation} In this simplified form, \(\chi\) can be seen to
represent the strength of the tail of each neighbour's spectrum after
transmission through the combined filter width \(\Delta \nu_c\).

Considering only the spectral crosstalk term, the resolution of the PTC
is given by Eq.~\ref{eq:app_enob_xt}. Setting the resolution to a target
\(B^*\), and solving for \(N f_B\) gives the spectral limit of
Sec.~\ref{sec:resolution_results}. With Eq.~\ref{eq:chi_beat}, this
inversion is exact and yields Eq.~\ref{eq:wall} with
\(k_\text{eff} = k_{xt}\) and
\(c_\text{eff} = \sqrt{a^2 + (\Delta\nu_c/f_B)^2} \approx 1.34\),
although the full form of Eq.~\ref{eq:chi} is always used in the
numerical evaluation. With the leaked intensity term included, there is
no closed form, so we solve for the limit numerically at each \(B^*\).
The form of Eq.~\ref{eq:wall} still describes these numerical limits to
within 0.5\% from 2 to 8 bits, with fitted values
\(c_\text{eff} = 1.42\) and \(k_\text{eff} = 0.79\). Compared with the
beat term alone, the leaked intensity term roughly doubles
\(k_\text{eff}\) and increases \(c_\text{eff}\) by about 6\%. It
decreases the crosstalk limits by 8--14\% against the beat term alone
(1.57 to 1.35 THz at 2 bits), but changes the efficiency of the most
efficient designs by less than 3\%, because the receiver and SOAs are
the dominant source of noise for most of the optimal designs.

The closed forms here hold only for Gaussian shaped PSD and filters, and
the numerical results depend strongly on these filter shapes. The
exponential tails of the Gaussian give rise to the soft \(\sqrt{\ln}\)
limit in Eq.~\ref{eq:wall}, while unshaped NRZ (PSD which decays as
\((f_B/\Delta\nu_{ch})^2\)) would yield a much harder limit,
\(N f_B \propto 1/\sqrt{\mathrm{SNR}^*} \propto 2^{-B^*}\). Nyquist
pulse shaping would allow for \(\Delta\nu_{ch} \approx (1 + \beta) f_B\)
with vanishing crosstalk. We choose Gaussian shapes as a tractable
middle ground, but the model should be evaluated numerically with the
measured PSD and filter responses for any real system design. The
derivation also only considers the two nearest neighbours, which is a
valid approximation for
\(\Delta\nu_{ch} \gtrsim 2\sqrt{\Delta\nu_{sig}^2 + \Delta\nu_c^2}\),
which should hold for any realistic channel separation.

\subsection{Link Budget and Optical
Amplification}\label{sec:link_budget}

\subsubsection{Passive losses}\label{passive-losses}

The total per-channel passive loss from source to detector is the sum of
all insertion loss terms and the fan-out loss,
\begin{widemath}\begin{equation}\protect\phantomsection\label{eq:link_loss}{
\mathrm{IL}_{tot}(N, S) = \mathrm{IL}_{mux} + 10\log_{10} N + 2 \mathrm{IL}_{mod} + \mathrm{IL}_{AWGR} + 10\log_{10} S + \mathrm{IL}_{demux},
}\end{equation}\end{widemath}\noindent  where
\(\mathrm{IL}_{mod} = \mathrm{IL}_{MZM} + \mathrm{IL}_{quad} + \mathrm{IL}_{cpl}\)
is the loss per modulator stage, comprising the MZM's insertion loss,
the 3 dB quadrature-bias transmission, and, in the heterogeneously
coupled scenario, the chip coupling loss per stage. The
\(10\log_{10} N\) term is the fan-out loss required to broadcast the
laser signals to all \(N\) modulators, and \(10\log_{10} S\) likewise
for the spatial fan-out after the AWGR. The total loss is equal to a
fixed overhead of order 15--25 dB, plus a scaling
\(\mathrm{IL}_{tot} \propto 10\log_{10}(NS)\), which means that any
large-scale instance of this architecture must include amplification to
compensate, and we must be careful about how we accumulate signal gain
and noise depending on the amplifier placement in the link.

\subsubsection{ASE accumulation and the cascade noise
figure}\label{ase-accumulation-and-the-cascade-noise-figure}

Each SOA with gain \(G\) and noise figure \(F\) adds single-polarization
ASE with spectral density \(S_{ASE,0} = n_{sp}(G-1)h\nu\), where the
spontaneous-emission factor \(n_{sp}\) follows from the exact relation
\(F = \tfrac{1}{G}(2 n_{sp}(G-1) + 1)\), and is around
\(n_{sp} \approx F/2\) at high gain. For a given placement of an
amplifier in the link, the ASE it injects experiences all the downstream
loss and gain, and so the total spectral density of the ASE arriving at
the detector must be treated explicitly using the forward recursion
relation,
\begin{equation}\protect\phantomsection\label{eq:ase_recursion}{
\begin{aligned}
S_{ASE} &\to g\, S_{ASE} \quad \text{(through any element with gain/loss } g\text{)}, \\ S_{ASE} &\to S_{ASE} + S_{ASE,0} \quad \text{(at each amplifier output)},
\end{aligned}
}\end{equation} where \(S_{ASE}\) is initialized to zero at the source.
Through this propagation formula, every subsequent amplifier multiplies
the ASE power accumulated up until that point and then adds its own
contribution, while loss decreases the accumulated ASE power. After
summing over the link, this gives
\(S_{ASE} = \sum_k S_{ASE,0,k}\, t_k\), where \(t_k\) is the net
transmission from the output of amplifier \(k\) to the detector.

The recursion is exact for any placement and, in the high-gain limit, it
reduces to the Friis cascade formula for the end-to-end noise figure,
\(F_{sys} = F_1 + \sum_{k\ge 2} (F_k - 1)/\prod_{j<k} g_j\), where
passive loss stages with transmission \(g_k\) are equivalent to
amplifiers with noise figure \(1/g_k\). Passive loss adds nothing to
Eq.~\ref{eq:ase_recursion}, its noise figure simply expresses that the
signal falls relative to the shot noise, which the receiver model counts
through the detected power {[}38{]}. The classical results of the Friis
equation hold, where a gain stage before a loss stage preserves the
noise figure \(F_{sys} \approx F_{SOA}\), while loss before gain
multiplies it to \(F_{sys} \approx F_{SOA}/g_k\). Amplifier placement
therefore directly determines \(P_{det}\) and \(S_{ASE}\) at the
detector.

\subsubsection{Saturation and the placement
constraint}\label{saturation-and-the-placement-constraint}

Amplifiers have four different possible placement locations in this
architecture, each of which has a different multiplicity in the device
count. They can be placed before the fan-out to the first modulator bank
(\(\times 1\)), after the first modulator bank (\(\times N\)), after the
second modulator bank (\(\times NS\)), or after the demultiplexer
(\(\times N^2 S\)). A choice of placement defines the integer vector
\((M_0, M_1, M_2, M_3)\) which represents the number of gain stages at
each possible placement location, with total device count
\(N_{SOA} = \sum_i n_i M_i\), where \(n_i \in \{1, N, NS, N^2S\}\).
Placing them as far upstream as possible is desirable, due to the
reduced multiplicity in the device count, and due to the Friis results
described in the previous section, but this is constrained by
limitations on the optical power, described below.

The SOAs have a saturation limit to their aggregate output optical
power, which we implement as \(n_\lambda P_{ch} G \le P_{sat}\), where
\(n_\lambda\) is the number of WDM lines (\(n_\lambda = N\) in most
cases except after the demultiplexer where \(n_\lambda = 1\)). The other
optical power limits we impose are a cap at any point in the waveguides
\(P_{wg}\) which is set to limit the system to the linear optical
regime, which avoids distortion components which would be introduced
with higher power, and a detector saturation cap
\(P_{det} \le P_{det,max}\) which ensures linear operation at the
detector. In the TFLN scenarios, the multiplexer, the AWGR and the
routing are silicon nitride, which carries much more power than lithium
niobate, so the model applies \(P_{wg,max}\) to the silicon nitride and
a lower cap \(P_{wg,mod}\) to the aggregate power at the input of each
modulator bank (Tbl.~\ref{tbl:hardware_params}). Because all \(N\) lines
share the multiplexer output, the source power per line is also capped
at \(\min(P_{0,max}, P_{wg,max}/N)\). The SOA saturation power is the
most limiting exactly where the SOA would be most desirable, since a
single shared SOA before splitting the pump input waveguide would serve
all channels with only a single device, but its per-channel output is
capped at \(P_{sat}/N\), which decreases as more channels are used.
Increasing \(N\) and \(S\) therefore forces gain stages downstream of
all the insertion and splitting losses, and increases the minimum
achievable noise figure \(F_{sys}\) (directly related to the ENOB in the
regime where the ASE noise dominates). This is a fundamental scaling
relationship for this architecture, as quantified in
Sec.~\ref{sec:results}.

\subsubsection{Analysis objective}\label{analysis-objective}

The model is evaluated via a search over every design point (\(N\),
\(S\), \(f_B\)). At each point, the search solves for the per-line
source power \(P_0\), the amplifier placement (\(M_0\ldots M_3\)) and
the gain of each stage, and Eq.~\ref{eq:ase_recursion},
Eq.~\ref{eq:master_snr} and Sec.~\ref{sec:power_model} are used to
compute the ENOB and the total electrical power of each candidate
design. We then use three different selection criteria over this search
depending on the analysis. To evaluate the achievable resolution after
accounting for the link, the results of Fig.~\ref{fig:link_limited} take
the candidates with the highest ENOB. To survey the most efficient
points in the design space, Fig.~\ref{fig:design_space} and the
efficiency frontier of Fig.~\ref{fig:pareto} and
Tbl.~\ref{tbl:frontier_designs} take, for each target resolution
\(B^*\), the candidate design with the lowest total power that reaches
\(B \ge B^*\), with the data converter power chosen for \(B^*\). The
throughput frontier of Fig.~\ref{fig:pareto} takes, for each target
resolution \(B^*\), the candidate design with the highest throughput
among those that reach \(B \ge B^*\).

The algorithm for determining laser power and SOA placement is not an
exact optimization; instead, we implement a heuristic search over the
space according to rules that should yield optimal designs. The laser
power \(P_0\) is swept over 16 logarithmically spaced values over the 30
dB below its limit. For the SOAs, each of the listed SOA slots can hold
up to 3 stages of gain, where each stage takes the largest gain, up to
\(G_{max}\), that the saturation and waveguide caps allow, plus an extra
limit to the final slot to respect the limit \(P_{det}\). After the SOA
stages are filled, we reduce the gain of the final stage between 0 and
12 dB in steps of 2 dB, which can save amplifier power when the result
is already above the target resolution. To limit the search, we skip SOA
placements when the total gain of the placement lies outside the range
\([\mathrm{IL}_{tot} - 45, \mathrm{IL}_{tot} + 10 + G_{max}]\) dB. The
design space we consider is \(N \in \{1, \ldots, 64\}\),
\(S \in \{1, 2, 4, \dots, 64\} \cup \{N\}\), and 32 log-spaced symbol
rates \(f_B\) between 2 GBaud and \(f_{B,max}\). Every efficiency result
in Sec.~\ref{sec:results} is the optimum this algorithm found within
this sampled space. As a convergence check on the most efficient
designs, we repeated the same search around the three best points with a
finer grid (\(N \pm 2\), every integer \(S\) from \(S/2\) to \(2S\), 15
rates within a factor of 1.5, 32 source powers and up to 4 stages per
slot), which raised their efficiency by at most 0.6\%. Compared to the
frontier results in Fig.~\ref{fig:pareto}, the plots in
Fig.~\ref{fig:design_space} use a finer grid, with every integer \(N\)
up to 64, every integer \(S\) up to 16 plus
\(S \in \{20, 24, 32, 48, 64\}\), and 51 symbol rates from 2 to 200
GBaud.

\subsection{Electrical Power Model}\label{sec:power_model}

\subsubsection{Laser sources}\label{laser-sources}

For DFB lasers, the electrical power per line is
\begin{equation}{P_L = V_L(P_0/\eta_L + I_{th}),}\end{equation} where
\(\eta_L\) is the slope efficiency in W/A, \(I_{th}\) is the threshold
current, and \(V_L\) the bias voltage {[}42{]}. In addition to the raw
electrical power, lasers require thermoelectric stabilization to
maintain stability of their output wavelength, which we model using a
thermoelectric cooler (TEC). If the TEC power is equal to twice the heat
generated, \(P_L - P_0\), then the total power of \(N\) laser sources is
equal to, \begin{equation}\protect\phantomsection\label{eq:laser_power}{
P_{lasers} = N(P_L + 2(P_L-P_0)).
}\end{equation} A frequency-comb source replaces the array with a single
pump at different (typically higher) per-line RIN.

\subsubsection{Modulators and drivers}\label{modulators-and-drivers}

Each travelling-wave MZM dissipates RF power \(V_{RMS}^2/R_L\) into a
matched termination load \(R_L=50~\Omega\), and also requires thermal
power, on average half of a thermal \(\pi\)-shift, for maintaining a
quadrature bias point, so \begin{equation}{
P_{MZM} = V_{RMS}^2/R_L + \tfrac{1}{2}P_{\pi,th}.
}\end{equation} O-band TFLN modulators at 100 GBaud have been shown to
use around \(P_{MZM} \approx 33\) mW {[}23{]}. We take this value in
this work, although we are performing our analysis in the C-band.
Because \(V_\pi\) scales roughly with wavelength, a C-band device needs
about 1.2 times the drive voltage and 1.4 times the RF power, which may
raise the total power by a few percent at the designs of
Sec.~\ref{sec:frontier}.

We estimate that the power of a driver amplifier is approximately linear
in the symbol rate and quadratic in the size of the drive swing, so we
use the equation, \begin{equation}{
P_{drv} = P_{drv,ref}\,\frac{f_B}{f_{ref}} \left(\frac{V_{pp}}{V_{ref}}\right)^2,
}\end{equation} where we anchor our estimate with \(P_{drv,ref} = 100\)
mW at \(f_{ref} = 112\) GBaud, which we take from the specifications of
a co-packaged CMOS driver {[}43{]}. That work reports only the switching
current, not the swing voltage, but we estimate from the published
specifications a reference voltage of \(V_{ref} \approx 2\) V
peak-to-peak. The swing needed for each modulator is given by its
half-wave voltage \(V_{pp} = (2/\pi)\arcsin(m)\,V_\pi\), which is simply
\(V_\pi\) when \(m = 1\).

The total power consumption for all the modulators including their
drivers is then, \begin{equation}{
P_{mod} = (N + SN)(P_{MZM} + P_{drv}).
}\end{equation}

\subsubsection{Data converters}\label{data-converters}

Because one of the central goals of this work is to investigate the
efficiency of the PTC against its resolution and its speed, we must
explicitly use a model for the data converter power which incorporates
its resolution \(B_{conv}\) and sample rate \(f_{Sa}\). We use the
Walden-style {[}44{]} model,
\begin{equation}\protect\phantomsection\label{eq:converter}{
P_{DAC/ADC} = \mathrm{FoM} \cdot 2^{B_{conv}} \cdot f_{Sa},
}\end{equation} where the figure of merit \(\mathrm{FoM}\) is a commonly
used metric to evaluate state of the art data converters. From surveys
of ADC and DAC performance and figures of merit {[}45, 46{]}, we take
\(\mathrm{FoM} = 40\) fJ/conversion-step for both ADCs and DACs, near
the high-speed envelope of the survey data. The system requires
\(N + SN\) DACs at \(f_{Sa} = 2f_B\) and resolution \(B_{conv} = B^* +\)
margin. The ADCs require a slightly different treatment here here
because of the accumulation over symbols. Because an \(L\)-symbol analog
accumulation leaves the per-receiver readout rate at \(f_B/L\) but
requires \(\tfrac{1}{2}\log_2 L\) additional output bits, conversion
power amortizes as \(1/\sqrt{L}\),
\begin{equation}\protect\phantomsection\label{eq:adc_power}{
P_{ADC} = N^2 S\cdot \mathrm{FoM}\cdot 2^{B_{conv}}\cdot\frac{f_B}{\sqrt{L}},
}\end{equation} with the target converter resolution
\(B_{conv} = B^* + \text{margin}\) (the \(\tfrac{1}{2}\log_2 L\) extra
output bits are carried by the \(\sqrt{L}\) factor). The results take
\(L = N\), so the data converter term scales as \(N^{3/2}S\), although
in principle the PTC can be operated with \(L\gg N\), which can drive
down the converter power contribution. Time-multiplexing one ADC across
\(N\) wavelength channels of a spatial output reduces the device count
from \(N^2S\) to \(SN\) but it maintains the overall power consumption
unchanged when \(N\approx L\). Workloads with a large inner dimension
\(L \gg N\) amortize the power as \(\sqrt{N/L}\).

The total converter power is then given by \begin{equation}{
P_{conv} = (N + SN) P_{DAC} + SN\, P_{ADC}.
}\end{equation}

\subsubsection{Receivers}\label{receivers}

Each of the \(N^2 S\) receivers uses a photodiode with negligible power
consumption (we model \(P_{PD} \approx 1\) mW), and a TIA integrator.
The integrator is the bulk of the receiver power, and depends on the
bandwidth. Just as for the modulator drivers, we anchor our estimate
from a measured reference, which reports a power of \(P_{TIA,ref}=11.2\)
mW at \(f_{ref}=64\) GBaud {[}47{]}. We then approximate a linear power
scaling, \begin{equation}{
P_{TIA} = P_{TIA,ref} \frac{f_B}{f_{ref}},
}\end{equation} where the integrating configuration is assumed to draw
no extra power, and the TIA noise density appears in
Sec.~\ref{sec:receiver_model}. The scaling of power with bandwidth for
TIAs is less well characterized than for converters, which is a source
of uncertainty in the current model.

\subsubsection{Optical amplifiers}\label{optical-amplifiers}

Each SOA stage in our model is characterized by a maximum gain
\(G_{max}\), its noise figure \(F\), and an aggregate output saturation
power \(P_{sat}\). The total power consumption of the device is just the
electrical power \(P_{SOA} = V_{bias} I_{bias}\) at full gain. In
general, these specifications can often trade off against each other,
because output saturation power grows roughly proportionally with the
bias current. In our model, we take all four values from published
demonstrations for each of the platforms we consider.

We allow for the gain \(G\) of each SOA to be varied between 0 dB and
\(G_{max}\), and we model the power consumption as varying linearly with
\(G\), from zero up to \(P_{SOA}\). This results in a total amplifier
power model, \begin{equation}{
P_{amps} = P_{SOA} \sum_k n_k\, \frac{G_k^{dB}}{G_{max}^{dB}}
}\end{equation} where the sum is over all placement locations and
\(n_k\) is the number of SOAs with gain \(G_k^{dB}\) placed at location
\(k\). We keep the noise figure and saturation power constant as gain is
varied, which isn't accurate for real SOAs, since they require a nonzero
bias current before they become transparent and begin to amplify, and
saturation power falls as bias current increases.

For the hybrid TFLN platform, we use the quantum-dot SOA (QD SOA) of
St-Arnault et al., with maximum gain of 18.2 dB, a 5.6 dB noise figure,
and a 3-dB output saturation power of 19.4 dBm at 1.6 V bias drawing 700
mA of current {[}48{]}. For the silicon platform, we use a heterogeneous
III--V-on-silicon SOA with 25 dB of gain, 13.7 dBm saturation power,
which draws 0.41 W {[}49{]}. We choose the QD SOA as the reference even
though it has a higher power consumption because the saturation power
limits the received power in every design we investigated, so the
relevant figure of merit for an amplifier in this architecture is the
output power it can deliver before saturating per Watt of bias, rather
than its power per device. The same study measured a quantum-well and a
bulk SOA with similar gain and noise figure, a saturation power of only
about 8 dBm, and about 0.35 W per stage. With such a device (8 dBm, 0.35
W), the most efficient TFLN designs are 1.1--1.3 times less efficient at
2 and 3 bits, 1.3 times less efficient at 4 bits and 2.0 times less
efficient at 5 and 6 bits, so the quantum-dot device is the best of the
measured set.

\subsubsection{Summary}\label{summary}

The total power consumption of the system is the sum of all these listed
contributions,
\(P_{total} = P_{lasers} + P_{mod} + P_{conv} + P_{rx} + P_{amps}\), and
then the computational efficiency \(\eta\) of the PTC, in units of
OPS/W, is given by,
\begin{equation}\protect\phantomsection\label{eq:ptotal}{
\eta = \frac{T}{P_{total}} = \frac{2N^2S f_B}{P_{total}}.
}\end{equation}

\subsection{Technology platform values}\label{sec:instantiation}

We chose component values for the numerical evaluation of the model to
represent achievable values of two current technology platforms so that
the results represent currently realizable performance of the PTC
architecture. The first is a monolithic silicon-on-insulator platform,
where for the lasers and SOAs we take values for a heterogeneously
integrated III-V platform, which we use as a point of comparison to the
TFLN platform, which is the baseline case we investigate. The TFLN
baseline represents a system with TFLN modulators bonded to a silicon
nitride chip which holds the WDM components---the multiplexers, the
AWGR, and the demultiplexers---and III-V SOA chips for the gain, which
are coupled at about 2 dB per facet. Using silicon nitride for the WDM
components allows those sections to carry up to 30 dBm, but the
aggregate power in the TFLN platform is limited to 20 dBm. Coupling loss
between the modulator platform and the gain medium cannot be avoided, so
we make sure to represent it in the model. As another comparison to the
TFLN with similar parameters, rather than a completely different set, we
also evaluate a ``multi-chip'' TFLN scenario, in which the modulator,
AWGR, multiplexer/demultiplexer, and SOA chips are all coupled
separately, with more pessimistic coupling losses, closer to what an
experimental demonstration in a lab might achieve. In addition to the
specific values chosen for each platform, we take into account another
platform-specific mechanism to make sure we represent physically
meaningful potential technologies, which is a validity region for the
achievable symbol rate on each platform (55 GBaud for monolithic silicon
and 200 GBaud for hybrid TFLN), so that we don't inadvertently arrive at
results which couldn't be possible in reality.
Tbl.~\ref{tbl:hardware_params} lists all the physical parameters used in
the numerical instantiation of the model that generated all the results
for the different platforms, while Tbl.~\ref{tbl:shared_params} gives
parameters shared by all evaluations of the model. For the comparison of
the published Pappas results in Sec.~\ref{sec:discussion}, we use a 4 dB
loss for the multiplexer and demultiplexer, 5 dB for the AWGR, 4 dB for
the modulators, EDFA-class amplifiers with 21 dBm saturation, 5 dB noise
figure and 5 W of power consumption, a 30 dBm waveguide cap, 0.65 A/W
responsivity and a 10 dBm power cap for the photodetectors, and 16
wavelengths with a 100 GHz channel spacing.

\begin{table*}[tbp]
\centering
\caption{Hardware parameters for the three technology scenarios. The TFLN
platforms use SiN passives and LN modulators. Multi-chip values have the
sources of the heterogeneous TFLN column, with the losses of extra
facets. \(^\dagger\)Assumed. \(^\ddagger\)Estimated.}\label{tbl:hardware_params}
{\def\LTcaptype{} 
\begin{tabular}{@{}
  >{\raggedright\arraybackslash}p{(\linewidth - 6\tabcolsep) * \real{0.2791}}
  >{\raggedright\arraybackslash}p{(\linewidth - 6\tabcolsep) * \real{0.2558}}
  >{\raggedright\arraybackslash}p{(\linewidth - 6\tabcolsep) * \real{0.2558}}
  >{\raggedright\arraybackslash}p{(\linewidth - 6\tabcolsep) * \real{0.2093}}@{}}
\toprule\noalign{}
\begin{minipage}[b]{\linewidth}\raggedright
Parameter
\end{minipage} & \begin{minipage}[b]{\linewidth}\raggedright
Si + het. III--V
\end{minipage} & \begin{minipage}[b]{\linewidth}\raggedright
TFLN het.
\end{minipage} & \begin{minipage}[b]{\linewidth}\raggedright
TFLN multi-chip
\end{minipage} \\
\midrule\noalign{}
\(f_{B,max}\) {[}GBaud{]} & 55 {[}43{]} & 200 {[}50{]} & 200 \\
\(\mathrm{IL}_{mux}\)/\(\mathrm{IL}_{demux}\) {[}dB{]} & 1.5/2.0
{[}51{]} & 1.5/2.0 {[}52{]} & 1.5/3.0 \\
\(\mathrm{IL}_{MZM}\) {[}dB{]} & 3.5 {[}43{]} & 1.0 {[}50, 53{]} &
1.5 \\
\(V_\pi\) {[}V{]} & 7.6 {[}43{]} & 2.3 {[}50{]} & 2.3 \\
\(\mathrm{IL}_{cpl}\) {[}dB/modulator{]} & 0.3\(^\dagger\) & 1.0 {[}53,
54{]} & 2.5 {[}22, 52{]} \\
\(\mathrm{IL}_{SOA,cpl}\) {[}dB/SOA{]} & 2.0 {[}55{]} & 4.0 {[}56{]} &
5.0 {[}57{]} \\
\(\mathrm{IL}_{AWGR}\) {[}dB{]} & 2.7 {[}58{]} & 3.0 {[}59{]} & 4.5 \\
\(G_{max}\) {[}dB{]} & 25 {[}49{]} & 18.2 {[}48{]} & 18.2 \\
\(P_{sat}\) {[}dBm{]} & 13.7 {[}49{]} & 19.4 {[}48{]} & 19.4 \\
\(P_{wg,max}\)/\(P_{wg,mod}\) {[}dBm{]} & 17/17\(^\ddagger\) {[}60,
61{]} & 30 {[}62{]}/20 {[}63{]} & 30/20 \\
\(NF\) {[}dB{]} & 8 {[}49{]} & 5.6 {[}48{]} & 5.6 \\
\(P_{SOA}\) {[}W{]} & 0.41 {[}49{]} & 1.12 {[}48{]} & 1.12 \\
\bottomrule
\end{tabular}
}
\end{table*}

\begin{table*}[tbp]
\centering
\caption{Parameters shared by all platforms. \(^\dagger\)Assumed.}\label{tbl:shared_params}
{\def\LTcaptype{} 
\begin{tabular}{@{}
  >{\raggedright\arraybackslash}p{(\linewidth - 4\tabcolsep) * \real{0.2000}}
  >{\raggedright\arraybackslash}p{(\linewidth - 4\tabcolsep) * \real{0.4000}}
  >{\raggedright\arraybackslash}p{(\linewidth - 4\tabcolsep) * \real{0.4000}}@{}}
\toprule\noalign{}
\begin{minipage}[b]{\linewidth}\raggedright
Group
\end{minipage} & \begin{minipage}[b]{\linewidth}\raggedright
Parameter
\end{minipage} & \begin{minipage}[b]{\linewidth}\raggedright
Value
\end{minipage} \\
\midrule\noalign{}
Spectrum & \(\Delta\nu_{tot}\) & 4 THz {[}38{]} \\
& \(a\), \(a_o\) & 1, 2\(^\dagger\) \\
& \(\Delta f_{el}\), \(\Delta f\) & \(f_B\)\(^\dagger\), \(f_B/2\) \\
& \(\Delta\nu_{o,min}\), \(\Delta\nu_{ch,min}\) & 20 GHz, 25
GHz\(^\dagger\) \\
& \(\kappa\) & 1.5\(^\dagger\) (App. A) \\
Signal Source & \(m\), \(\sigma_{op}^2\), \(\sigma_d^2\) & 1, 1/3,
1/9 \\
& RIN & \(-155\) dB/Hz {[}25, 64{]} \\
& \(P_{0,max}\) & 13 dBm {[}64, 65{]} \\
& \(\eta_L\), \(I_{th}\), \(V_L\) & 0.3 W/A, 10 mA, 1.8 V {[}42{]} \\
Modulators & \(\mathrm{IL}_{quad}\) & 3 dB \\
& \(P_{MZM}\) & 33 mW {[}23{]} \\
& \(P_{drv,ref}\) at \(f_{ref}\) & 100 mW at 112 GBaud {[}43{]} \\
Receiver & \(\mathcal{R}\) & 0.8 A/W {[}66, 67{]} \\
& \(\bar{\imath}_{TIA}\) & 18 pA/\(\sqrt{\mathrm{Hz}}\) {[}47{]} \\
& \(P_{TIA,ref}\) at \(f_{ref}\) & 11.2 mW at 64 GBaud {[}47{]} \\
& \(P_{PD}\) & 1 mW\(^\dagger\) \\
& \(P_{det,max}\) & 3 dBm {[}68{]} \\
Converters & FoM & 40 fJ/conv.-step {[}45, 46{]} \\
& \(B_{conv} - B^*\) & 1 bit\(^\dagger\) \\
& \(L\) & \(N\)\(^\dagger\) \\
\bottomrule
\end{tabular}
}
\end{table*}

\subsection{Model scope}\label{sec:model_scope}

We have deferred the study of a number of physical effects in this
analysis to future work. Most of these effects would reduce ENOB or
efficiency, so the current results constitute upper bounds in the
context of the following changes.

\emph{Nonlinearity.} We assume a linear IM/DD link for all platforms. We
also assume perfect \(\arcsin\) predistortion in
Sec.~\ref{sec:receiver_model}, without considering the power and latency
cost of this operation, but real predistortion wouldn't be perfectly
linear. The MZM drivers and the photoreceivers are taken to be linear
and don't introduce any intersymbol interference. Distortion and ISI
would require changing the resolution metric from SNR to SNDR {[}69,
70{]}, but the analysis would otherwise remain the same. All of these
effects grow with the symbol rate, which agrees with the gap between our
model and measurements.

\emph{Amplifier model limitations.} The SOA saturation and waveguide
power limits count the carrier power only, but modulation peaks, and
effects such as cross-gain modulation and the buildup of ASE power would
all contribute towards these limits, which are already the limiting
effects in many cases. For variable gain, we estimated that SOA power is
linear in its dB gain for simplicity, but this is an incorrect model. An
improved model would take the dependence of gain, saturation power, and
noise figure on bias current into account, including a nonzero
transparency current. Another flaw is that the QD SOA values come from
O-band measurements, while the rest of the model uses the C-band.

\emph{Noise and crosstalk.} The in-band crosstalk between AWGR ports is
the most important unmodelled term, as discussed in
Sec.~\ref{sec:discussion}, but can also in principle be calibrated out.
Another source of optical noise is how the RIN of a comb source would
differ from the multiplexed single mode lasers we consider. The optical
crosstalk also assumes channels with the same polarization, and a
Gaussian spectrum with margin \(\kappa=1.5\). The ENOB values only hold
exactly under these conditions and with uniform operands.

\emph{Component power.} The AWGR loss and isolation in a real AWGR
should depend on its port count, but we could not find published
analyses on what this scaling relationship should be. The loss should
increase with \(N\), which would push the optimal designs toward lower
channel counts. We also don't count power consumption for calibration
and modulator bias control, the analog multiplexers of the shared ADC
readout, or gain and loss variation across the band.

\emph{System design and application.} We only consider the components of
the system up to the boundary between the analog compute and the digital
interface. Memory, I/O, and thermal management beyond the TEC and MZM
bias heaters are unmodelled. The results also assume \(L=N\), which is
an implicit limit on the workload of the PTC. A larger \(L\) depth would
reduce the converter power consumption.

\section*{Data and Code Availability}\label{data-and-code-availability}

The code needed to reproduce the figures and data in this analysis is
available in the Github repository available at the following URL:
https://github.com/Shastri-Lab/hypermultiplexed-ptc-scaling-analysis

\section*{Acknowledgments}\label{acknowledgments}

This work was supported in part by the Natural Sciences and Engineering
Research Council of Canada (NSERC), Canada Foundation for Innovation
(CFI) and Ontario Research Fund. B.J.S. is supported by the Canada
Research Chairs Program and the Sloan Research Fellowship.

\section*{Author Contributions}\label{author-contributions}

H.M, N.A-K, A.A, and B.J.S conceived of the original idea for the
scaling analysis. H.M, N.A-K, and A.A wrote an early version of the
analysis. H.M investigated all the dependencies between parameters to
extend the analysis, wrote the code for the simulations, and generated
the figures. H.M then wrote the main manuscript text with input from
N.A-K and A.A. B.J.S supervised this work.

\section*{References}\label{references}

\protect\phantomsection\label{refs}

\appendix

\section{Spectral crosstalk derivation}\label{sec:app_xtalk}

To evaluate the spectral crosstalk, we consider a victim channel \(i\)
with carrier frequency \(\nu_i\). With channel separation
\(\Delta\nu_{ch}=\Delta\nu_{tot}/N\), an aggressor channel \(j\) sits at
frequency \(\nu_j = \nu_i + (j - i)\,\Delta\nu_{ch}\). We take only the
two neighbouring channels to be aggressors, so the offset between a
victim and its two aggressor channels is \(\pm\Delta\nu_{ch}\). From
Eq.~\ref{eq:cascaded_mzm}, the intensity of channel \(j\) is given by,
\begin{equation}\protect\phantomsection\label{eq:app_mu}{
p_j(t) = P\,\big[1 + \mu_j(t)\big], \qquad \mu_j = m\,w_j + m\,x_j + m^2\,w_j x_j,
}\end{equation} where \(P\) is carrier power. We assume that both
\(x_j\) and \(w_j\) have zero mean and are independent, uncorrelated
signals with variance \(\sigma_{op}^2\). The product \(w_j x_j\) then
also has zero mean and has variance \(\sigma_d^2=\sigma_{op}^4\). The
variance of the signal on channel \(j\) is then
\begin{equation}\protect\phantomsection\label{eq:app_var}{
\sigma_\mu^2 = 2 m^2 \sigma_{op}^2 + m^4 \sigma_d^2.
}\end{equation} If the operands \(x\) and \(w\) have uniform
distributions and \(m=1\), then \(\sigma_{op}^2 = 1/3\),
\(\sigma_d^2 = 1/9\), and \(\sigma_\mu^2 = 7/9\). The linear terms make
up around 86\% of the signal variance that enters as crosstalk, and in
principle much of this could be calibrated out since the weight terms
are known beforehand, but we assume the full signal contributes to the
noise.

The field associated with the signal in Eq.~\ref{eq:app_mu} is
\begin{equation}\protect\phantomsection\label{eq:app_field_exact}{
E_j(t) = \sqrt{p_j(t)}\; e^{i(2\pi\nu_j t + \phi_j)} \approx \sqrt{P}\,\Big[1 + \tfrac{1}{2}\mu_j(t)\Big]\, e^{i(2\pi\nu_j t + \phi_j)},
}\end{equation} which contains a carrier term with power \(P\) and a
sideband term with power \(\tfrac{1}{4} P \sigma_\mu^2\). The PSD of the
sideband field can then be written as
\(\tfrac{1}{4} P \sigma_\mu^2\,\Phi(f - \nu_j)\), where \(\Phi\) is the
shape of the PSD. For this analysis, we take the PSD to have a Gaussian
profile, \begin{equation}\protect\phantomsection\label{eq:app_psd}{
\Phi(f) = \frac{1}{\sqrt{\pi}\,\Delta\nu_{sig}} \exp\!\left(-\frac{f^2}{\Delta\nu_{sig}^2}\right),
}\end{equation} where the \(1/e\) half-width \(\Delta\nu_{sig} = a f_B\)
and \(a \approx 1\). Any other \(\Phi\) (the \(\mathrm{sinc}^2\) of NRZ,
a raised cosine, etc.) could be evaluated numerically with the same
approach, but assuming Gaussian shape allows for easier algebraic
manipulation.

The WDM fields are demultiplexed with a filter which passes the field
with a power response \(H_o(f)\), which we also take to be Gaussian,
\begin{equation}\protect\phantomsection\label{eq:app_ho}{
H_o(f) = \exp\!\left(-\frac{f^2}{\Delta\nu_o^2}\right),
}\end{equation} with \(1/e\) width
\(\Delta\nu_o=\max\!\big(a_o\Delta\nu_{sig},\, \Delta\nu_{o,min}\big)\),
where the field response is \(h_o(f)=H_o(f)^{1/2}\). The insertion loss
of the multiplexer is already taken into account in the link budget, so
we are free to set \(H_o(0) = 1\). Similarly, the electrical filter of
the receiver is taken to be Gaussian, with a power response,
\begin{equation}{
H_{el}(f) = \exp\!\left(-\frac{f^2}{\Delta f_{el}^2}\right),
}\end{equation} where \(\Delta f_{el} = f_B\).

The field at the detector is then given by
\(E = E_i + \sum_{j \ne i} \tilde{E}_j\), where \(\tilde{E}_j\) is the
filtered field of neighbour \(j\),
\begin{equation}\protect\phantomsection\label{eq:app_leak}{
\tilde{E}_j(f) = h_o(f)\, E_j(f),
}\end{equation} and we assume that the victim field passes its own
filter with no additional attenuation or distortion.

The response at the receiver with responsivity \(\mathcal{R}\) is then
given by, \begin{equation}\protect\phantomsection\label{eq:app_terms}{
\frac{i}{\mathcal{R}} = \underbrace{|E_i|^2}_{\text{wanted}} + \underbrace{2\sum_{j\neq i}\mathrm{Re}\big[E_i \tilde{E}_j^*\big]}_{\text{beat}} + \underbrace{\sum_{j\neq i}|\tilde{E}_j|^2}_{\text{leaked intensity}},
}\end{equation} where we ignore the negligible contribution from the
two-neighbour term.

\subsection{Beat term}\label{beat-term}

The beat current between the carrier with power \(P\) and a stationary
random field with PSD \(S(f)\) has a variance given by
\(2\mathcal{R}^2 P \int S(f)\,df\). The component of the beat current
coming from the sideband of neighbour \(j\) with an offset \(f\) from
carrier \(i\), beats with a frequency \(|f|\), so it passes through both
filters with an attenuation \(H_o(f) H_{el}(f)\). For one of these
neighbours, this results in a contribution to the noise current variance
for channel \(i\) given by,
\begin{equation}\protect\phantomsection\label{eq:app_beat}{
\langle i_{beat,j}^2 \rangle = I_{ph}^2 \frac{\sigma_\mu^2}{2} \int H_o(f)\, H_{el}(f)\, \Phi(f \mp \Delta\nu_{ch})\, df,
}\end{equation} where the combined response of the filters is
\(H_o(f)\, H_{el}(f) = \exp(-f^2/\Delta\nu_c^2)\), with
\(\Delta\nu_c^{-2} = \Delta\nu_o^{-2} + \Delta f_{el}^{-2}\). Evaluating
the integral in Eq.~\ref{eq:app_beat}, using the definition of
Eq.~\ref{eq:app_psd}, yields,
\begin{equation}\protect\phantomsection\label{eq:app_overlap}{
\int \Phi(f - \Delta\nu_{ch})\, \exp\!\left(-\frac{f^2}{\Delta\nu_c^2}\right)\, df = g' \exp\!\left(-\frac{\Delta\nu_{ch}^2}{\Delta\nu_{sig}^2 + \Delta\nu_c^2}\right),
}\end{equation} where
\(g' = \Delta\nu_c/\sqrt{\Delta\nu_c^2 + \Delta\nu_{sig}^2}\) is the
fraction of the sideband power which passes through the combined filter
at zero detuning. The total noise current variance from both neighbours
is then given by,
\begin{equation}\protect\phantomsection\label{eq:app_beatsum}{
\langle i_{beat}^2 \rangle = g'\, \sigma_\mu^2 \, I_{ph}^2 \exp\!\left(-\frac{\Delta\nu_{ch}^2}{\Delta\nu_{sig}^2 + \Delta\nu_c^2}\right).
}\end{equation}

\subsection{Leaked intensity term}\label{leaked-intensity-term}

The leaked intensity term is the beat between the leaked carrier of
neighbour \(j\) and its own leaked sidebands, so it does not depend on
the victim. The leaked carrier has field amplitude
\(\sqrt{P}\,h_o(\Delta\nu_{ch})\), and its sideband at an offset
\(\delta\) from \(\nu_j\) is attenuated by
\(h_o(\Delta\nu_{ch} + \delta)\). Unlike the beat term, the carrier and
its sidebands come from the laser so they have a fixed phase offset.
Since \(\mu_j\) is real, the sidebands at \(\pm \delta\) are a conjugate
pair, and they add in amplitude rather than in power when they beat with
the carrier. The two sidebands pass through the victim's demultiplexer
with different amplitudes and lead to a noise current variance of,
\begin{widemath}\begin{equation}\protect\phantomsection\label{eq:app_li}{
\langle i_{leak,j}^2 \rangle = I_{ph}^2 \frac{\sigma_\mu^2}{4}\, H_o(\Delta\nu_{ch}) \int \Phi(\delta)\,\big[h_o(\Delta\nu_{ch} + \delta) + h_o(\Delta\nu_{ch} - \delta)\big]^2\, H_{el}(\delta)\, d\delta.
}\end{equation}\end{widemath}\noindent 

For Gaussian filters,
\(h_o(\Delta\nu_{ch} + \delta)\,h_o(\Delta\nu_{ch} - \delta) = H_o(\Delta\nu_{ch})\,H_o(\delta)\),
and we can evaluate the integrals in the same way as was done for
Eq.~\ref{eq:app_overlap}, which yields
\begin{widemath}\begin{equation}\protect\phantomsection\label{eq:app_li_gauss}{
\langle i_{leak,j}^2 \rangle = I_{ph}^2 \frac{\sigma_\mu^2}{2}\, g' \left[\exp\!\left(-\frac{\Delta\nu_{ch}^2}{\Delta\nu_o^2}\right)\, \exp\!\left(-\frac{\Delta\nu_{ch}^2}{\Delta\nu_o^2 + \Delta\nu'^2}\right) + \exp\!\left(-2\frac{\Delta\nu_{ch}^2}{\Delta\nu_o^2}\right)\right],
}\end{equation}\end{widemath}\noindent  where
\(\Delta\nu'^{-2} = \Delta\nu_{sig}^{-2} + \Delta f_{el}^{-2}\) is the
combined width of the signal and the electrical filter, and \(g'\) is
the same factor as in Eq.~\ref{eq:app_overlap}.

\subsection{Neglected terms and field
nonlinearity}\label{neglected-terms-and-field-nonlinearity}

Several terms in Eq.~\ref{eq:app_terms} are neglected by our model. The
victim carrier also beats with the leaked neighbour carrier, which
produces a tone at \(\Delta\nu_{ch}\) which carries no data, but has a
variance of \(2 H_o(\Delta\nu_{ch}) H_{el}(\Delta\nu_{ch})\, I_{ph}^2\)
per neighbour. If the lasers are phase-locked (as in a frequency comb),
this tone would be a fixed waveform which could be calibrated out, but
would add an extra noise term in the case of independent lasers. Since
the tone is at a frequency \(\Delta\nu_{ch}\), it can also be designed
to be cut off by the electrical filter of the receiver. The victim's own
sidebands also beat with the leaked carrier of the neighbour, which adds
a copy of the victim's own data shifted by a frequency
\(\Delta \nu_{ch}\) before the filter of the receiver, which is only a
small contribution and is further attenuated by the receiver. The
remaining terms are all either constant, of higher order in \(\mu\), or
contribute beat terms at frequencies well beyond the pass band of the
filters. This derivation also only holds for victims with two or more
neighbours, so it is not exact in the case of \(N \le 2\).

There is a small correction that we model due to the higher order terms
that are dropped from Eq.~\ref{eq:app_field_exact}. The term on the
order of \(\mu^2\) has spectral shape given by \(\Phi * \Phi\), which is
up to twice as wide as \(\Phi\), so it adds power to the far tails of
the neighbour's spectrum, where it contributes to the beat term. To
estimate the size of this effect, we ran a small numerical simulation of
the exact field \(\sqrt{(1 + m w)(1 + m x)}\) of one neighbour, with
\(w\) and \(x\) being drawn uniformly from \([-1, 1]\) for each symbol,
and Gaussian pulse shaping. We then filtered the resulting field with
\(H_o\), computed the beat term with the victim carrier, and applied the
electrical filter shape \(H_{el}\), the result of which is compared with
the variance of Eq.~\ref{eq:app_beat}. We find that the simulated beat
is 0.4 to 1.1 times the first-order result for \(\Delta\nu_{ch} = 2\) to
\(5 f_B\), which supports the first order result, but there remains a
source of uncertainty due to the unknown shape of the PSD of the
modulation. To account for this uncertainty, we include a margin factor
of \(\kappa=1.5\) incorporated into the beat term.

\subsection{Total crosstalk and spectral
limit}\label{total-crosstalk-and-spectral-limit}

Adding the beat term of eq.~\ref{eq:app_beatsum}, with the margin factor
\(\kappa\), to the leaked intensity term of Eq.~\ref{eq:app_li_gauss},
the total crosstalk noise can be modelled as,
\begin{equation}\protect\phantomsection\label{eq:app_kxt}{
\langle i_{xt}^2 \rangle = k_{xt}\,\chi\, I_{ph}^2,
}\end{equation} where
\(k_{xt} = \tfrac{1}{2}\,\sigma_\mu^2\, g'\, \kappa\) is a coefficient
that represents the strength of the beat term, and
\begin{widemath}\begin{equation}\protect\phantomsection\label{eq:chi}{
\chi = 2 \exp\!\left(-\frac{\Delta\nu_{ch}^2}{\Delta\nu_{sig}^2 + \Delta\nu_c^2}\right)
+ \frac{2}{\kappa}\left[\exp\!\left(-\frac{\Delta\nu_{ch}^2}{\Delta\nu_o^2} - \frac{\Delta\nu_{ch}^2}{\Delta\nu_o^2 + \Delta\nu'^2}\right) + \exp\!\left(-\frac{2\Delta\nu_{ch}^2}{\Delta\nu_o^2}\right)\right],
}\end{equation}\end{widemath}\noindent  is the leakage factor that
measures how much the channel spacing affects the strength of the
crosstalk. The first term of Eq.~\ref{eq:chi} represents the beat term,
and the second represents the leaked intensity.

Considering Eq.~\ref{eq:app_kxt} as the only source of noise, and
evaluating the bit resolution using Eq.~\ref{eq:enob} and a full-scale
current \(I_\text{FS} = 2m^2 I_{ph}\), we find that,
\begin{equation}\protect\phantomsection\label{eq:app_enob_xt}{
4^{B} = \frac{I_\text{FS}^2}{12\,\sigma_n^2} = \frac{m^4}{3\,k_{xt}\,\chi}.
}\end{equation} Keeping only the beat term of Eq.~\ref{eq:chi}, then
\(\chi = 2\exp(-\Delta\nu_{ch}^2/C^2)\) with
\(C^2 = \Delta\nu_{sig}^2 + \Delta\nu_c^2\), and we can solve for the
channel spacing at a target resolution \(B^*\),
\begin{equation}\protect\phantomsection\label{eq:app_wall_step}{
\Delta\nu_{ch}^2 = C^2 \left[2\ln(2)\,B^* + \ln\!\big(6 k_{xt}/m^4\big)\right].
}\end{equation} Substituting \(N = \Delta\nu_{tot}/\Delta\nu_{ch}\)
gives Eq.~\ref{eq:wall}, with \(k_\text{eff} = k_{xt}\) and
\(c_\text{eff} = C/f_B\). With the leaked intensity included there is no
closed form, so we solve for the limit numerically in
Fig.~\ref{fig:wall_master}.

\end{document}